\PassOptionsToPackage{letterpaper,margin=0.96in}{geometry}
\documentclass[pdflatex,sn-nature]{sn-jnl}

\usepackage{graphicx}%
\usepackage{multirow}%
\usepackage{amsmath,amssymb,amsfonts}%
\usepackage{amsthm}%
\usepackage{mathrsfs}%
\usepackage[title]{appendix}%
\usepackage{xcolor}%
\usepackage{textcomp}%
\usepackage{manyfoot}%
\usepackage{booktabs}%
\usepackage{algorithm}%
\usepackage{algorithmicx}%
\usepackage{algpseudocode}%
\usepackage{listings}%

\usepackage{algorithm}
\usepackage{algpseudocode}
\usepackage{array}      
\usepackage{tabularx}   
\usepackage{makecell}   
\newcolumntype{Y}{>{\centering\arraybackslash}X} 
\usepackage{threeparttable}

\theoremstyle{thmstyleone}%
\theoremstyle{thmstyletwo}%

\theoremstyle{thmstylethree}%

\begin{document}

\title[Article Title]{LIMO: \underline{L}ow-Power \underline{I}n-Memory-Annealer and \underline{M}atrix-Multiplication Primitive for Edge C\underline{o}mputing}


\author*[1]{\fnm{Amod} \sur{Holla}}\email{aholla@purdue.edu}

\author[2]{\fnm{Sumedh} \sur{Chatterjee}}\email{ee22b005@smail.iitm.ac.in}
\equalcont{These authors contributed equally to this work.}

\author[3]{\fnm{Sutanu} \sur{Sen}}\email{sutanu2004@kgpian.iitkgp.ac.in}
\equalcont{These authors contributed equally to this work.}

\author[1]{\fnm{Anushka} \sur{Mukherjee}}\email{mukher96@purdue.edu}

\author[4]{\fnm{Fernando} \sur{Garc\'{i}a-Redondo}}\email{Fernando.GarciaRedondo@imec-int.com}

\author[5]{\fnm{Dwaipayan} \sur{Biswas}}\email{dwaipayan.biswas@imec.be}

\author[6, 1]{\fnm{Francesca} \sur{Iacopi}}\email{Francesca.Iacopi@imec-int.com}

\author*[1]{\fnm{Kaushik} \sur{Roy}}\email{kaushik@purdue.edu}

\affil*[1]{\orgdiv{Elmore School of Electrical and Computer Engineering}, \orgname{Purdue University}, \orgaddress{\city{West Lafayette}, \postcode{47907}, \state{Indiana}, \country{USA}}}

\affil[2]{\orgdiv{Department of Electrical Engineering}, \orgname{Indian Institute of Technology}, \orgaddress{\city{Chennai}, \postcode{603103}, \state{Tamil Nadu}, \country{India}}}

\affil[3]{\orgdiv{Department Of Electronics And Electrical Communication Engineering}, \orgname{Indian Institute of Technology}, \orgaddress{\city{Kharagpur}, \postcode{721302}, \state{West Bengal}, \country{India}}}

\affil[4]{\orgname{imec}, \orgaddress{\city{Cambridge}, \country{UK}}}

\affil[5]{\orgname{imec}, \orgaddress{\city{Leuven}, \country{Belgium}}}

\affil[6]{\orgname{imec USA}, \orgaddress{\city{West Lafayette}, \postcode{47906}, \state{Indiana}, \country{USA}}}

\abstract{Combinatorial optimization (CO) underpins critical applications in science and engineering, ranging from logistics to electronic design automation. A classic example of CO is the NP-complete Traveling Salesman Problem (TSP). Finding exact solutions for large-scale TSP instances remains computationally intractable; on von Neumann architectures, such solvers are constrained by the memory wall, incurring compute–memory traffic that grows with instance size. Metaheuristics, such as simulated annealing implemented on compute-in-memory (CiM) architectures, offer a way to mitigate the von Neumann bottleneck. This is accomplished by performing in-memory optimization cycles to rapidly find approximate solutions for TSP instances. Yet this approach suffers from degrading solution quality as instance size increases, owing to inefficient state-space exploration. To address this, we present LIMO, a programmable mixed-signal computational macro that implements an in-memory annealing algorithm with reduced search-space complexity. The annealing process is aided by the stochastic switching of spin-transfer-torque magnetic-tunnel-junctions (STT-MTJs) to escape local minima. For large instances, our macro co-design is complemented by a refinement-based divide-and-conquer algorithm amenable to parallel optimization in a spatial architecture. Consequently, our system comprising several LIMO macros achieves superior solution quality and faster time-to-solution on instances up to 85,900 cities compared to prior hardware annealers. The modularity of our annealing peripherals allows the LIMO macro to be reused for other applications, such as vector-matrix multiplications (VMMs). This enables our architecture to support neural network inference. As an illustration, we show image classification and face detection with software-comparable accuracy, while achieving lower latency and energy consumption than baseline CiM architectures.}



\maketitle

\section{Introduction}\label{sec1}
\begin{figure}[h]
\centering
\includegraphics[width=0.73\textwidth]{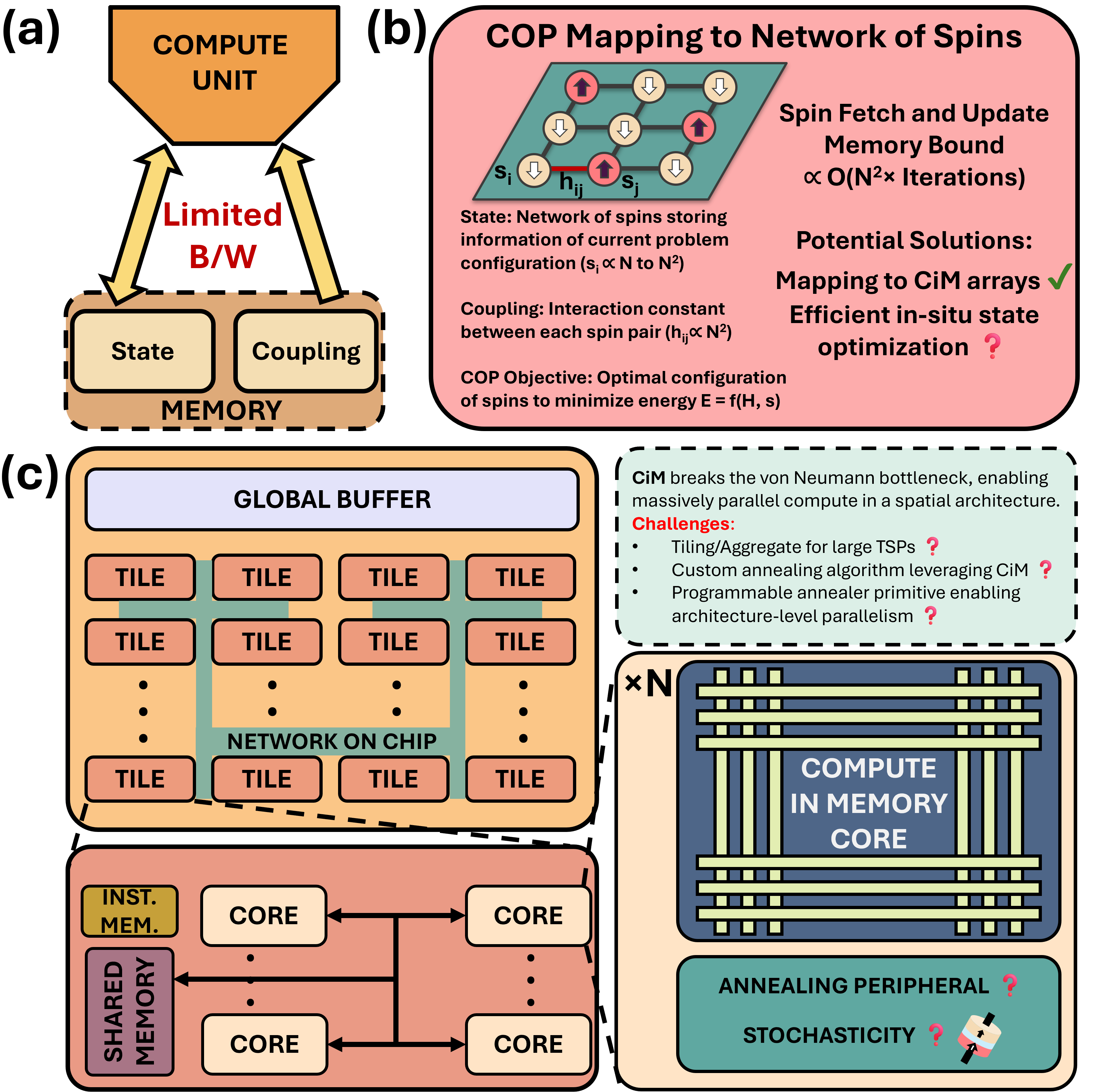}
\caption{(a) Illustration of the von Neumann bottleneck between memory and compute units for CO (b) Network-of-spins reformulation of COPs, with state access time growing quadratically. (c) A spatial architecture based on CiM primitives that need to efficiently solve divided CO instances. The considered architecture is organized hierarchically. Multiple tiles are connected by a network-on-chip. Each tile comprises multiple cores, and each core contains multiple compute units.}\label{fig0}
\end{figure}

Combinatorial optimization (CO) is a foundational class of problems in science and engineering, essential for applications ranging from logistics and vehicle routing to automated circuit design and genetic research \cite{melo2009facility, vehicle_routing, sechen1985timberwolf, wang1994msa}. A typical example of a CO problem (COP) is the Traveling Salesman Problem (TSP), which is NP-complete. This means that solving the TSP is computationally equivalent to solving other NP-complete COPs, such as 3-SAT or graph coloring, as they are reducible to one another in polynomial time \cite{GareyJohnson}. The TSP and its numerous variants have many direct applications in routing and scheduling \cite{Punnen2007Applications,Lawler1985TSP}. Consequently, the TSP serves as the representative COP in our study. The NP-hard nature of the TSP, like many other COPs, renders the exact solution of large-scale instances computationally intractable \cite{Applegate2006TSP}, even when employing state-of-the-art specialized solvers like Concorde \cite{Applegate2006TSP} or commercial Mixed Integer Linear Programming (MILP) solvers such as Gurobi \cite{Gurobi2024}. While heuristic methods like the Lin–Kernighan algorithm \cite{Helsgaun2000LKH} offer a more tractable, scalable alternative for finding near-optimal solutions to larger problems, they too become infeasible for instances exceeding several thousand cities. Metaheuristics such as simulated annealing (SA) provide rapid, high-quality tours in practice by seeking the global minima of the state-space, and stochastically accepting uphill moves to escape local minima \cite{simulated_annealing}. However, tour quality significantly degrades for large instances solved using standard SA owing to inefficient state-space exploration. Moreover, formal convergence to the global minima requires an impractically slow, logarithmic cooling schedule \cite{Granville1994Simulated, Hajek1988Cooling, Geman1984Stochastic}. Importantly, for TSP solvers implemented on conventional von Neumann architectures, performance is further limited by the memory wall, requiring repeated access to large state vectors and cost matrices, causing execution time to be dominated by data movement between compute and memory (Fig. \ref{fig0}(a\textendash b)) \cite{Wulf1995Memory}. Moreover, realizing robust on-chip stochasticity in the case of SA-based TSP solvers demands considerable silicon area and introduces computational latency \cite{Bakiri2018Survey, Fu2021Overview, Sunar2007Secure}. \\

Compute-in-memory (CiM) architectures can address the data-movement bottleneck by performing task-specific computations within memory arrays \cite{shafiee2016isaac}. This paradigm suits SA, and several CiM-based hardware implementations of annealers have been tailored for the TSP, mitigating the von Neumann bottleneck \cite{Si2024EnergyEfficient, yoo2025taxitravelingsalesmanproblem, Lu2024DigitalCIM}. These implementations generate on-chip stochasticity through various physical mechanisms, such as the telegraphic switching of low-barrier nanomagnets \cite{Si2024EnergyEfficient}, stochastic switching of spin-orbit-torque (SOT) devices \cite{yoo2025taxitravelingsalesmanproblem}, or noise in SRAM bit-cells induced by process variation \cite{Lu2024DigitalCIM}. While promising, these approaches are not without drawbacks. They typically require high-precision digital-to-analog converters to carefully control the annealing schedule, and their implementation of stochasticity in the analog domain compromises robustness. In addition, magnetic-memory based noise sources suffer from significant device-to-device variation \cite{mtj_rng1, mtj_rng2, smtj_variation}, while noisy-SRAM bit-cells are temporally constant sources of stochasticity, leading to correlated bit failures. Critically, implementations of these in-memory annealers have not proven scalable for TSPs, either being limited to small problem instances ($<15$ cities) \cite{Si2024EnergyEfficient} or exhibiting substantial tour-quality degradation at larger scales \cite{yoo2025taxitravelingsalesmanproblem, Lu2024DigitalCIM}. This degradation occurs because the hardware implementations cannot overcome the inefficient state-space exploration inherent to SA, a problem that is exacerbated as the problem size grows. Beyond these performance limitations, the specialized nature of these designs for annealing limits their utility, rendering them unsuitable for general-purpose memory or accelerator applications. For example, the low-barrier nanomagnet array in Si \emph{et al.} \cite{Si2024EnergyEfficient} suffers from poor data retention, while the SOT-MRAM crossbar arrays in Yoo \emph{et al.} \cite{yoo2025taxitravelingsalesmanproblem} are constrained by high write energies. \\This motivates the co-design of an efficient annealing algorithm with a programmable architecture that natively supports its parallel execution. This architecture should be based on a computational primitive that, while featuring modular and process-variation robust peripherals for annealing, is not limited to this single application and retains general-purpose utility. \\

To that effect, this article introduces LIMO, a hardware-algorithm co-designed computational primitive with an 8T-SRAM core that supports in-memory annealing for TSPs through modular peripherals. Our design integrates three principal contributions: a hardware-aware annealing algorithm that improves upon the performance of standard SA for TSPs; a stochastic module for annealing that is robust to variations, utilizing spin-transfer-torque magnetic tunnel junctions (STT-MTJs) as random number generators (RNGs); and a refinement-based divide-and-conquer strategy for large-scale TSPs, which leverages our primitives in a spatial accelerator architecture.
Our annealing algorithm reduces the sample space selection complexity from quadratic to linear per optimization cycle. This enables efficient solution-space exploration, increasing the probability of iterations that improve the tour quality. Empirically, we show that this yields superior TSP solutions compared to standard SA. The LIMO macro executes each optimization cycle of our algorithm entirely in-situ. We incorporate an array of foundry-compatible STT-MTJs to supply the required stochasticity for annealing, leveraging their stochastic switching behavior. The stochasticity is implemented by following an analog-to-digital approach -- in contrast to digital-to-analog sources in prior works -- allowing each MTJ to function as an independent stochastic-bit source while most optimization operations occur in the digital domain. Our mixed-signal implementation, therefore, captures the benefits of both domains: the reliability of digital logic for the optimization operations and the energy-efficient stochastic switching of STT-MTJ devices in the analog domain. For very large-scale TSPs, we use a divide-and-conquer strategy augmented by refinement iterations that leverages a spatial architecture \cite{puma} composed of multiple LIMO primitives. This approach enables parallelized tour optimization by partitioning a large instance into multiple sub-TSPs, each mapped and solved independently by our macro, followed by re-stitching the global tour. The modular nature of the peripherals required for annealing ensure that the core 8T-SRAM compute array retains general-purpose utility. As an illustration, we demonstrate that the primitive's design is amenable to quantized vector-matrix multiplications (VMMs), making neural network inference possible within the same accelerator architecture.

 Our simulations show that a LIMO macro consumes approximately 15 times lower power than the equivalent previous state-of-the-art TAXI \cite{yoo2025taxitravelingsalesmanproblem} macro during annealing. Further, our divide-and-conquer strategy demonstrates 37.5\% tour quality improvement on average over the previous art across instances from the TSPLib benchmark suite. Cycle-accurate simulations of the spatial architecture (Fig. \ref{fig0}(c)) show a fivefold speedup over TAXI \cite{yoo2025taxitravelingsalesmanproblem} for solving an 85,900-city TSP, with superior solution quality. Concurrently, leveraging the macro's VMM capabilities through hardware-aware training, our system achieves comparable accuracy to software baselines on image classification and face detection workloads. The absence of analog-to-digital converters makes our design $\sim1.3-2.1\times$ more energy-efficient and $\sim1.2-1.3\times$ faster for serving CNN inference compared to baseline CiM architectures.

\section{Results}\label{sec2}
 Section \ref{subsec: SWAI} introduces our modified annealing algorithm that is used at the macro level. Section \ref{subsec: hw_sw_cod} presents the hardware-algorithm co-design of the LIMO macro; Section \ref{subsubsec: anneal_mode} describes the annealing mode that implements the stochastic optimizer for TSPs, and Section \ref{subsubsec: vmm_mode} details the operation of the VMM mode within the same compute-in-memory core. Section \ref{subsec: dnc_large_tsp} develops the divide-and-conquer strategy used to solve large-scale TSPs. Section \ref{subsec: adcless_algo} elucidates the hardware-aware algorithm for training neural networks that implement VMMs using our macro as a primitive. Section \ref{subsec: circuit sims} reports circuit-level simulations and energy/latency for both modes. Sections \ref{subsec: sys_sim_tsp} and \ref{subsec: sys_sim_cnn}  evaluate scaling via system-level simulations on arrays of LIMO macros in a spatial architecture based on PUMA \cite{puma}. The subsequent evaluation reports solution quality and time-to-solution for large-scale TSPs, in addition to inference accuracy, energy, and latency metrics for two representative CNN workloads.
\subsection{Significance Weighted Stochasticity in Annealing} \label{subsec: SWAI}
\begin{figure}[h]
\centering
\includegraphics[width=0.8\textwidth]{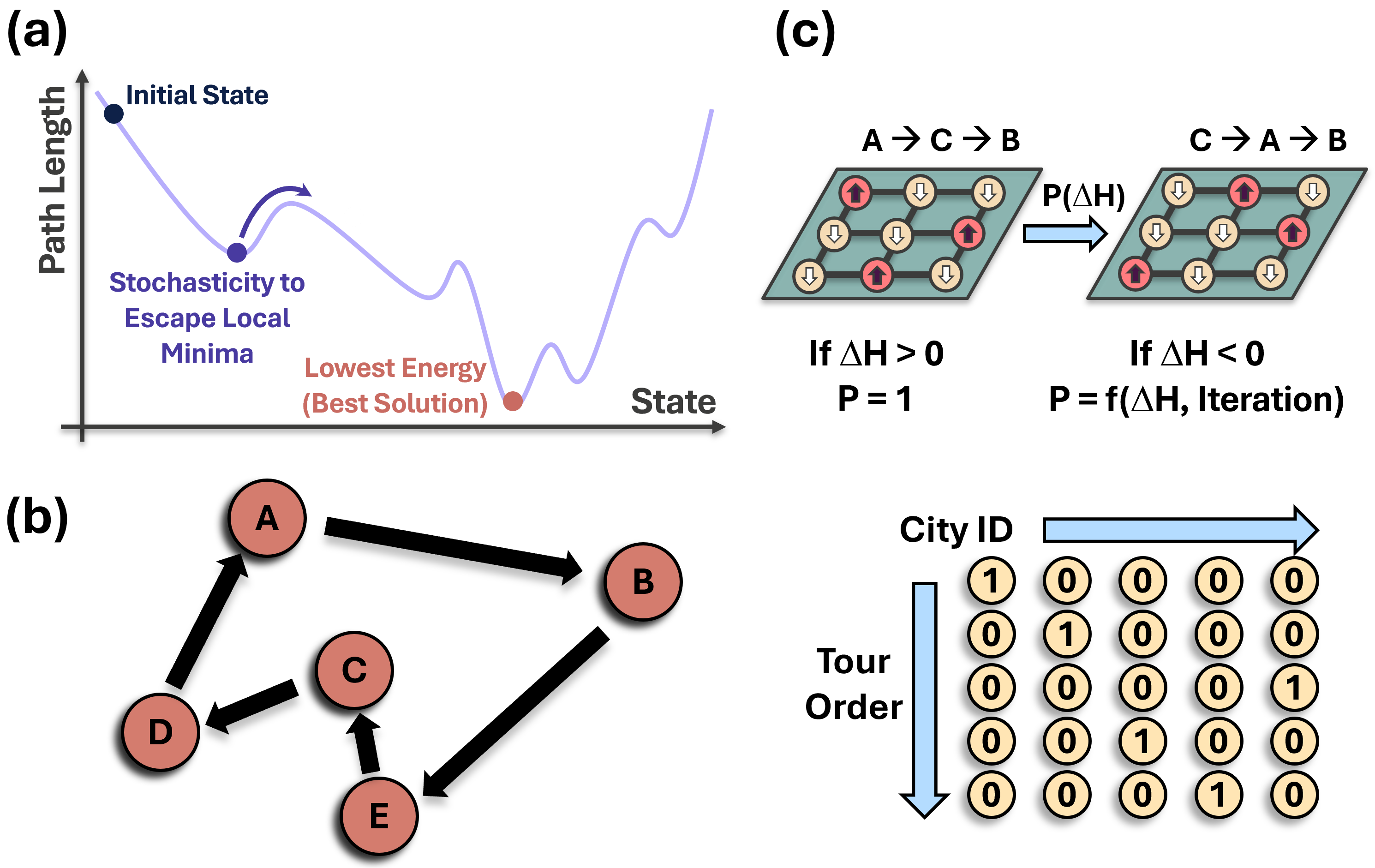}
\caption{ (a) Navigation of the energy landscape in simulated annealing. Each valid tour of cities has an associated tour length, analogous to an energy state. The simulated annealing algorithm iteratively modifies the current tour, primarily accepting changes that decrease the tour length. To escape local minima, the algorithm occasionally accepts energy-increasing moves. The probability of accepting these moves diminishes over the course of the optimization. (b) Mapping of a TSP to $N^2$ spins. Each row represents a tour position and is a one-hot vector representing a city that is visited at that tour position. (c) A state transition within the network-of-spins representation for the TSP. To maintain a valid tour, a move consists of swapping the columns corresponding to two cities, which is analogous to swapping their visiting order.}\label{fig1}
\end{figure}
\begin{figure}[h]
\centering
\includegraphics[width=0.95\textwidth]{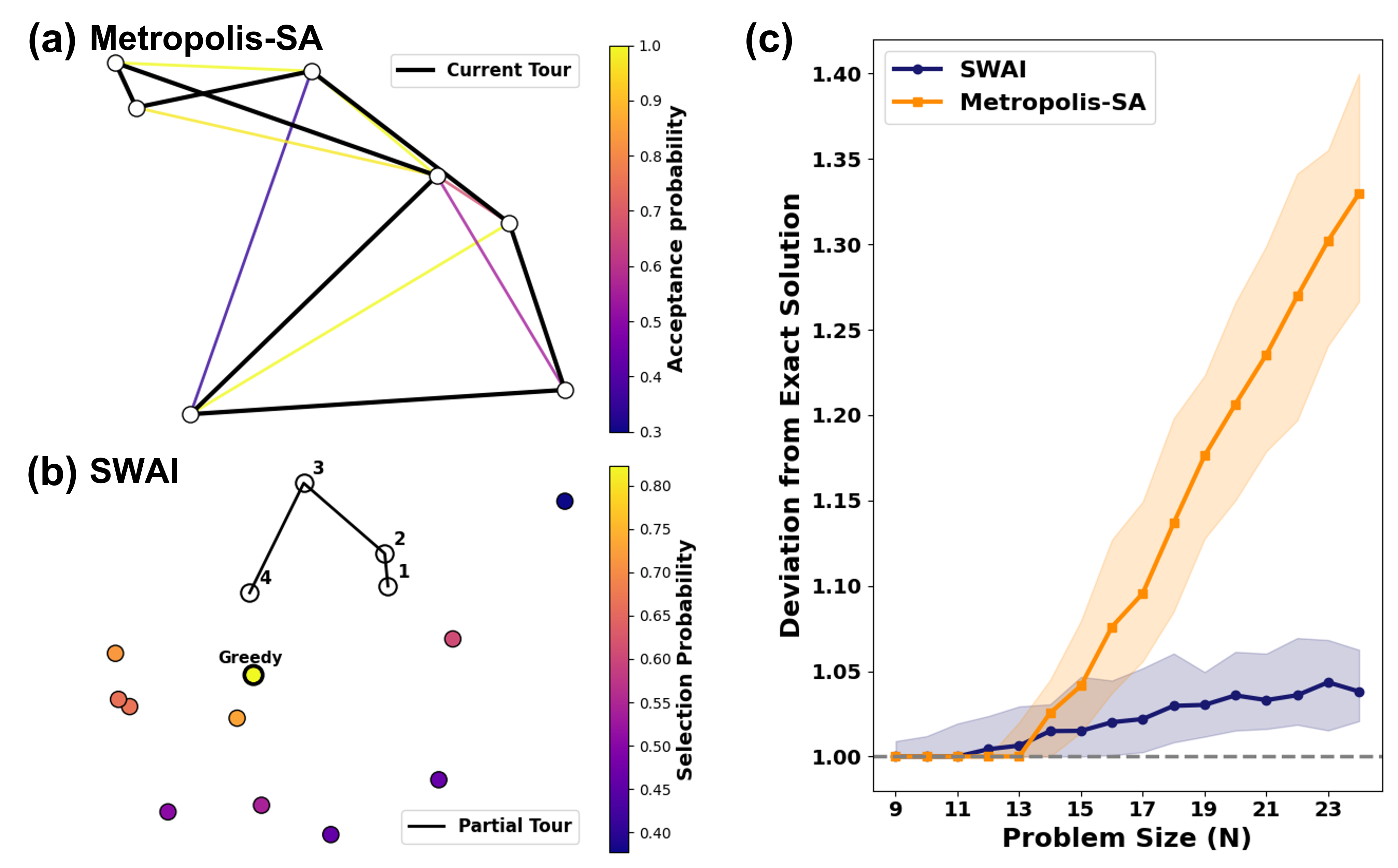}
\caption{(a) Iterative refinement process in Metropolis-SA. An edge heat-map is shown for a 7-city TSP instance that denotes the probability of accepting a swap between two cities if the pair is selected. Only edges with $P_{\text{accept}}>0.3$ are shown for clarity. (b) An illustration of the SWAI algorithm, that relies on iterative tour construction. A node heat-map is shown that denotes the selection probability. (c) Comparison of Metropolis-SA and SWAI on 500 random TSP instances per problem size, ranging from $N=9$ to $N=24$. Inter-quartile ranges are shaded around the median line-plots.}\label{fig2}
\end{figure}

For COPs like the TSP, SA serves as a versatile stochastic local search that efficiently navigates objective landscapes. SA starts with a random solution and, at each step, explores neighboring solutions (in the case of TSP, it examines the possibility of flipping the visiting order of two cities). It always accepts better solutions but will also sometimes accept worse ones, especially at the beginning of the optimization process (illustrated in Fig. \ref{fig1}(a)). This ability to accept worse solutions allows it to escape from local optima and explore more of the solution space. A TSP is usually mapped onto a densely-connected network of spins with each pair of spins described by a coupling constant (representing inter-city distances for a TSP). Each configuration of spins represents a certain solution in the solution space. Concretely, the Hamiltonian (or energy) for a system of \(N\) spins \(s = \{s_1, s_2, \dots, s_N\}\), with each \(s_i \in \{-1,1\}\), is
\begin{equation}
   H(s) \;=\; -\sum_{\langle i,j\rangle} J_{ij}\,s_i\,s_j \;-\; \sum_{i=1}^{N} h_i\,s_i.  
\end{equation}
 To map a TSP onto this Hamiltonian, the problem must be encoded in the coupling constants $J_{ij}$ and local fields $h_i$. The most common method uses $N^2$ spins ($N$ is the number of cities), where a spin $s_{v,p}$ represents whether city $v$ is at position $p$ in the tour (Fig. \ref{fig1}(b)) \cite{Lucas2014IsingFormulations}. The Hamiltonian is then constructed as a cost function that penalizes invalid tours (e.g., a city being in two positions at once, or a position being occupied by two cities) while rewarding short total distances. The objective is to minimize the total travel distance, which is encoded by setting the $J_{ij}$ values based on the distances between cities. A spin configuration that minimizes this Hamiltonian, is a valid, low-cost tour that can be considered to be the solution of the TSP.
The Metropolis acceptance rule \cite{Metropolis1953EoS} underpins classical SA.  
Given a current tour $\pi$ of length $H(\pi)$ and a candidate tour $\pi'$, the move
$\pi\!\to\!\pi'$ is accepted with probability  
\begin{equation}
  P_{\text{acc}}(\pi\!\to\!\pi') \;=\;
  \begin{cases}
    1, & \Delta H < 0,\\[4pt]
    \exp\!\bigl(-\tfrac{\Delta H}{T}\bigr), & \Delta H \ge 0,
  \end{cases}
  \qquad \Delta H = H(\pi')-H(\pi),
  \label{eq:metropolis}
\end{equation}
where $T$ is the temperature parameter that is gradually cooled over the optimization process till convergence (Fig. \ref{fig1}(c)).   
In the TSP, a move usually entails a pairwise swap of two cities (shown in Fig. \ref{fig2}(a)).  
As the problem size $N$ grows, swap–based SA tends to perform poorly because the search process through the rapidly growing state-space becomes increasingly inefficient. This is because the number of candidate edges to swap in the graph grows quadratically with $N$. \\
On a separate note, the Greedy Randomized Adaptive Search Procedure (GRASP)~\cite{Resende2003} employs biased randomized selection to \emph{iteratively construct} a tour from first to last city.  
During construction a restricted candidate list $\mathcal{R}$ is formed and an element $j\!\in\!\mathcal{R}$ is sampled with probability proportional to a bias function  
\begin{equation}
  g(j) = \bigl(f_{\max}-f(j)\bigr)^{\alpha}
\end{equation}
where $f(j)$ is the greedy score, $f_{\max}$ is the worst score in $\mathcal{R}$ and $\alpha>0$ is a bias-function parameter.  
Setting $\alpha=1$ and interpreting $f(j)$ as the edge distance $d_j$ yields the linear gate shown in equation \ref{eq:linear_gate}, which we subsequently adopt in our algorithm.
\begin{equation}
  P_j \;=\; 1 - \frac{d_j}{d_{\max}}
  \label{eq:linear_gate}
\end{equation}

We embed the linear gate given by equation \eqref{eq:linear_gate} inside an annealing schedule, while switching the move operator from swap to iterative insertion. Algorithm \ref{alg:SWAI} builds a tour left-to-right, repeatedly inserting a city at every iterated position. A ``pass'' is termed a single constructed tour after iterating from left to right. A city cannot be selected twice in a pass. 
At each tour position, a global Bernoulli variable $b\!\sim\!\mathrm{Bernoulli}(p)$ (termed as global stochastic-bit) is sampled that decides between pure greedy insertion (inserting the closest city to the one at the previous position) and stochastic insertion governed by the significance-weighted probability rule in equation \eqref{eq:linear_gate}. An illustration of algorithm \ref{alg:SWAI} is shown in Fig. \ref{fig2}(b). Similar to swap based simulated annealing, tour construction in our algorithm benefits from greedy insertion at late stages while still allowing uphill moves that enable solution space exploration.  
Unlike GRASP, which restarts multiple times with different bias functions and seeds, our method runs under a single annealing loop. The sample space selection complexity in our case, is therefore linear with respect to $N$ (compared to quadratic in the case of standard Metropolis-SA). \\

We evaluate the performance of Metropolis-SA against SWAI using a benchmark of randomly generated TSP instances. The benchmark consists of 500 instances for each problem size, ranging from 9 to 24 nodes, with the nodes for each instance being uniformly distributed across a unit square. We keep the number of iterations and cooling schedule identical between the two algorithms ($p_0=0.2$, $\beta=0.9995$, $p_{min}=0.01$). We solve each instance with an exact solver, and define a metric for solution quality given by equation \ref{eq:tsp_ratio}. 
\begin{equation}
    \label{eq:tsp_ratio}
    \text{Deviation Ratio} = \frac{\text{Tour Length given by Annealing Algorithm}}{\text{Tour Length given by Exact Solver}}
\end{equation}
A perfect solution implies a deviation ratio of 1. The empirical results in Fig. \ref{fig2}(c) demonstrate that SWAI scales significantly better than Metropolis-SA with respect to problem size. This improved scaling is likely because Metropolis-SA, which relies on pairwise swaps, explores the state space very locally at large $N$. Consequently, the probability of constructing long-range improvements becomes exponentially small, making SA prone to premature freezing with practical cooling schedules. In contrast, our algorithm's stochastic insertion iterations (an $\mathcal{O}(N)$ operation) enable more efficient exploration of the candidate space.

\begin{algorithm}[H]\
\label{alg:SWAI}
  \caption{Significance Weighted Annealed Insertion (SWAI)}
  \small
  \begin{algorithmic}[1]
    \Require Distance matrix $W\in\mathbb{R}^{N\times N}$, start city $s$, initial stochasticity $p_0$, decay factor $\beta$, cutoff probability $p_{min}$
    \State $\tau_{\text{best}}\gets[s]$\Comment{best tour found so far}
    \State $U\gets\{1,\dots,N\}\setminus\{s\}$\Comment{set of unused cities}
    \State $d_{\max}\gets\max(W)$\Comment{largest edge weight in $W$}
    \State $p\gets p_0$
    \While{$p\ge p_{\min}$} \Comment{annealing loop}
        \State $\tau\gets[s]$,\; $U'\gets U$ \Comment{fresh tour and unused set}
        \For{$k=2$ \textbf{to} $N$} \Comment{construct positions left--to--right}
            \State $\text{prev}\gets\tau[k{-}1]$ \label{alg_line: prev_sel}
            \State Draw $b\sim\mathrm{Bernoulli}(p)$ \label{alg_line: g_bit}\Comment{global stochastic bit}
            \If{$b=1$} \Comment{significance--weighted stochastic insertion}
                \State Compute $P_j\gets 1-W[\text{prev},j]/d_{\max}\quad\forall j\in U'$ \label{alg_line: local_gate}
                \State Select $j\in U'$ proportionally to $\{P_j\}$ \label{alg_line: selection_1}
            \Else \Comment{pure greedy insertion}
                \State $j\gets\arg\min_{u\in U'} W[\text{prev},u]$ \label{alg_line: selection_2}
            \EndIf
            \State Append $j$ to $\tau$;\; $U'\gets U'\setminus\{j\}$ \label{alg_line: insertion}
        \EndFor
        \If{$H(\tau)<H(\tau_{\text{best}})$} \Comment{update best solution}
            \State $\tau_{\text{best}}\gets\tau$ \label{alg_line: best_sol}
        \EndIf
        \State $p\gets\beta p$ \Comment{decay stochasticity}
    \EndWhile
    \State \Return $\tau_{\text{best}}$
  \end{algorithmic}
\end{algorithm}

\begin{figure}[h]
\centering
\includegraphics[width=\textwidth]{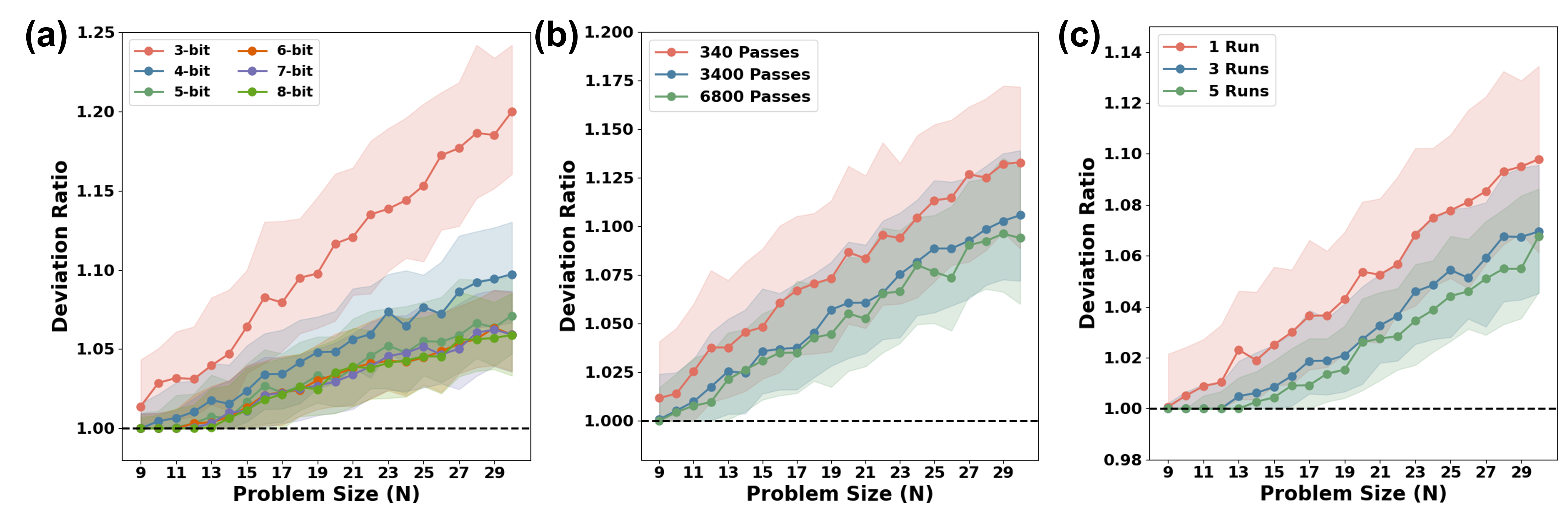}
\caption{Applying hardware related constraints to the SWAI algorithm. Inter-quartile ranges are shaded around the median line-plots. (a) Varying the bit-width of the weight matrix. (b) Varying the number of passes of the algorithm. (c) Varying the number of restart runs for a given problem instance.}\label{fig3}
\end{figure}
We perform a series of hardware-relevant ablation studies on the SWAI algorithm to concurrently investigate parameter influence on solution quality and derive target specifications for our proposed hardware. All tests in Fig. \ref{fig3} are conducted on 250 randomly generated TSP instances per problem size. The quality of the solution is adversely affected by weight matrix quantization, with a significant degradation observed for bit-widths of less than 4 (Fig. \ref{fig3}(a)). We further evaluate the impact of varying the number of passes (Fig. \ref{fig3}(b)) and restart runs (Fig. \ref{fig3}(c)). Increasing either parameter enhances solution space exploration and intuitively improves tour quality. This effect, however, empirically exhibits diminishing returns above approximately 3400 passes and 3 runs respectively.
\subsection{Hardware-Algorithm Co-Design of a LIMO Macro}
\label{subsec: hw_sw_cod}
\begin{figure}[h]
\centering
\includegraphics[width=\textwidth]{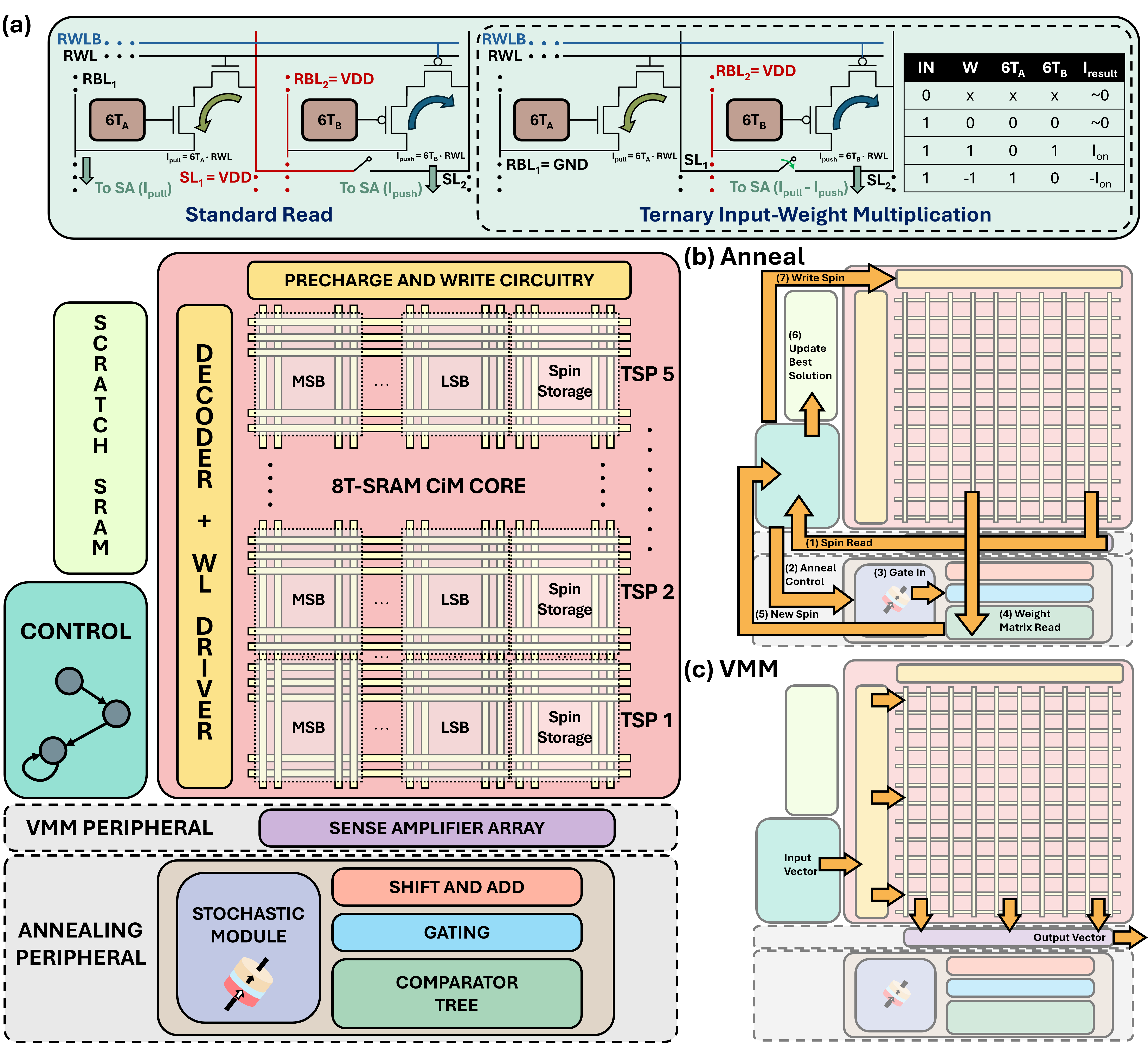}
\caption{Floorplan of the proposed LIMO macro. At its core is an 80$\times$80 8T-SRAM CiM crossbar that doubles as (i) the spin array with 4-bit programmable couplings for Ising annealing and (ii) a weight stationary matrix for neural-network inference. Distinct peripheral blocks are provisioned for the two operating modes; the annealing peripheral incorporates a compact STT-MRAM TRNG module. Five independent TSPs are concurrently mapped onto the crossbar and solved in parallel, while a scratch SRAM records the best tour for each instance. A finite-state-machine (FSM) based control module coordinates all dataflows between macro components across both modes. (a) Elementary bit-cell operations supported by the compute macro, including standard row-reads and ternary weight multiplication. Ternary weights are realized by a push-pull design in the read paths of neighboring bit-cells, which facilitates a bi-directional current flow to the source line. (b) Dataflow in annealing mode. (d) Dataflow in VMM mode.}\label{fig4}
\end{figure}

Building on the SWAI algorithm introduced in Section~\ref{subsec: SWAI}, we present the LIMO macro (floorplan illustrated in the main inset of Fig.~\ref{fig4}), which efficiently implements the algorithm's annealed insertion iterations while concurrently supporting VMMs for neural network inference under 1-bit partial-sum quantization. Following the approach in TAXI \cite{yoo2025taxitravelingsalesmanproblem}, we co-locate spins and their coupling weights within the same memory array to realise a highly efficient in-memory dataflow for the annealing process. To mitigate peripheral and TRNG overhead, the coupling matrix is limited to 4-bit precision. The macro comprises an 80$\times$80 CiM crossbar implemented in 8T-SRAM, with a modified bit-cell, shown in Fig. \ref{fig4}(a), where the read access paths for even columns comprise of PMOS transistors instead of NMOS. The rationale for this design choice is elaborated in section \ref{subsubsec: vmm_mode}. An $80\times80$ crossbar with a 4-bit coupling precision implies that our macro can support a maximum TSP size of 16. For annealing, this crossbar is logically partitioned into five independent 16$\times$80 sub-arrays, each dedicated to a single TSP instance. Within every 16$\times$80 sub-array, the left 16$\times$64 partition encodes the 4-bit coupling matrix (MSB $\ldots$ LSB), and the right 16$\times$16 partition stores the instantaneous tour configuration, also referred to as the spin storage. Two sets of peripherals exist for both modes. The annealing mode uses both sets of peripherals, while for VMMs, the output is taken from the sense amplifier array. A control block orchestrates the data-flow between all the components of the macro for both the modes. A scratch SRAM is included to record the best tour obtained in the annealing process for all 5 instances. Fig. \ref{fig4}(b) shows the macro-level dataflow in a single annealing iteration (operation explained in section \ref{subsubsec: anneal_mode}), while Fig. \ref{fig4}(c) illustrates the dataflow for obtaining 1-bit quantized VMMs (explained in section \ref{subsubsec: vmm_mode}). 

\subsection{Annealing Mode}
\label{subsubsec: anneal_mode}
The FSM controller executes every step of Algorithm~\ref{alg:SWAI} completely in-memory by coordinating the spin storage partition, the $4$-bit coupling matrix, and the peripheral logic, independently for each of the 5 problems. 
\\

\begin{figure}[h]
\centering
\includegraphics[width=\textwidth]{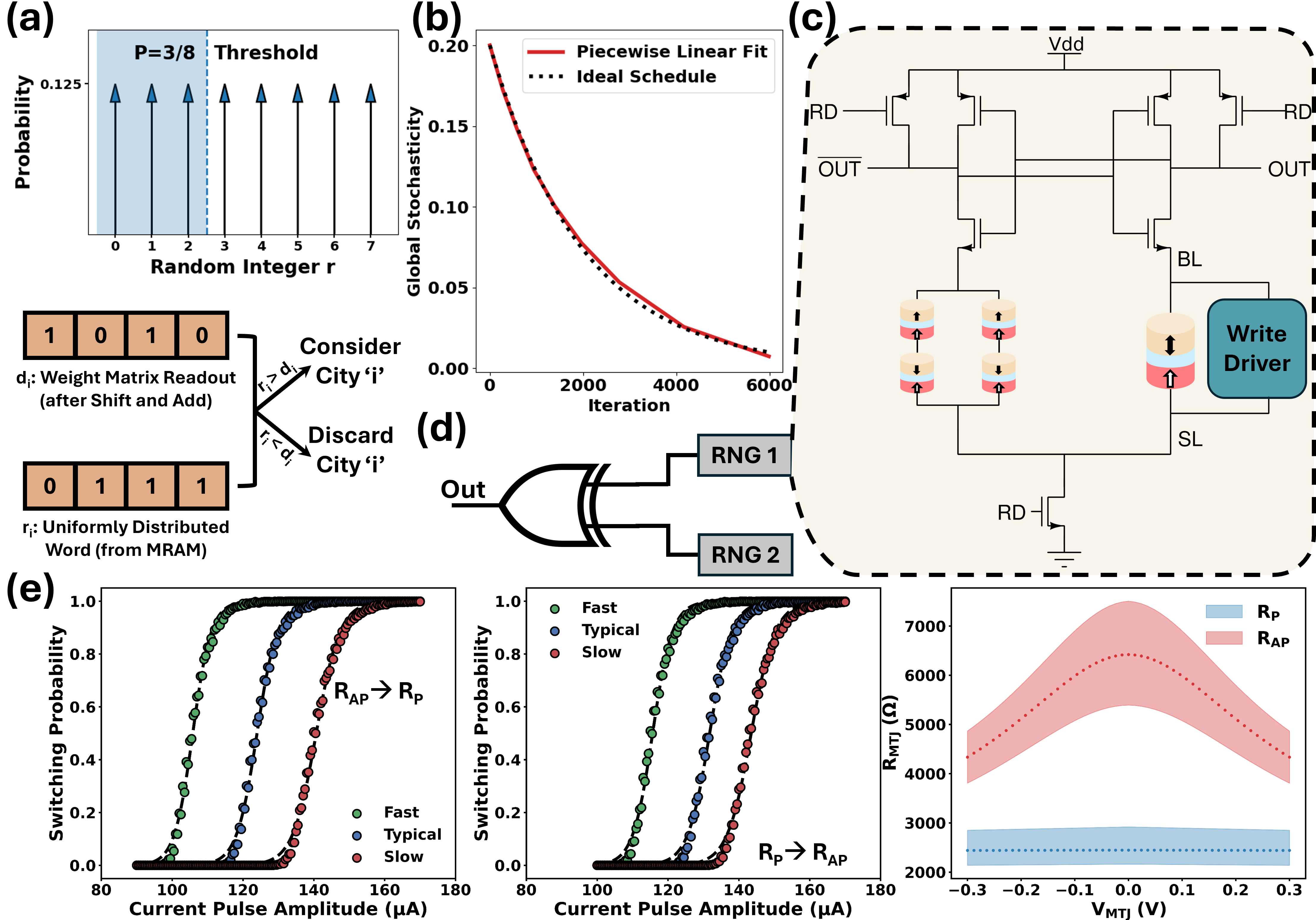}
\caption{(a) Illustrative example of the threshold based comparison to generate a Bernoulli sampler with a $3/8$ success probability from a 3-bit uniformly distributed number. This same concept is used to generate a stochastic gating vector that is dependent on the coupling constant value. (b) Ideal geometric decay with $p_0=0.2$ and $\beta=0.9995$ and its piecewise linear fit. (c) Circuit schematic of the sense‑amplifier‑based STT‑MRAM TRNG (d) XOR combination of the bitstreams from two identical instances yields a composite random sequence that is robust against device-to-device variations \cite{rhs-trng}. (e) Switching probability for the $R_{AP}\rightarrow R_{P}$ and $R_{P}\rightarrow R_{AP}$  transitions as a function of drive current (10ns pulse width) under fast, slow, and typical corners. The inset on the right plots the corresponding read resistance of the MTJ versus bias voltage for the same corners.}\label{fig5}
\end{figure}

Conventional hardware annealers described in literature typically generate an exponentially decaying annealing schedule through analog mechanisms. These approaches often exploit intrinsic device characteristics, such as the sigmoidal switching probability curve of MRAM \cite{yoo2025taxitravelingsalesmanproblem}, or modulate the supply voltage of SRAM elements to serve as a source of stochasticity \cite{Lu2024DigitalCIM}. Although theoretically elegant, such analog methods present significant practical implementation challenges. They necessitate fine-grained control over voltage sources, requiring high-precision digital-to-analog converters (DACs), and exhibit a high susceptibility to analog noise. \\
Our design implements stochasticity using a primarily digital methodology based on threshold comparison, that relies on comparing a uniformly distributed $N$-bit random integer, $r \in \{0, \dots, 2^N-1\}$, against a deterministic $N$-bit threshold, $d \in \{0, \dots, 2^N-1\}$. \\
Since each of the $2^N$ possible values of $r$ is equally likely, the predicate $[\,r < d\,]$ is \texttt{true} with a probability of $\Pr[r<d] = d / 2^{N}$.
Consequently, by encoding a desired success probability $p$ as the threshold $d = \lfloor p \cdot 2^{N} \rfloor$, we can realize a Bernoulli sampler with a success rate of approximately $p$ at the cost of only a single digital comparator. This concept is illustrated in Fig. \ref{fig5}(a). \\

We apply this principle at two distinct levels to control the annealing process: globally and locally.
First, a \emph{global} stochastic-enable bit, $s_{\mathrm g}$ (line \ref{alg_line: g_bit} of the algorithm), is generated by comparing a $16$-bit uniformly distributed random word, $r_{\mathrm g}$, against a reference word, $r_{\mathrm{ref}}$, supplied by the controller.
This comparison yields the predicate $s_{\mathrm g} = \bigl[r_{\mathrm g} < r_{\mathrm{ref}}\bigr]$.
The FSM effectively controls the annealing schedule by progressively lowering the value of $r_{\mathrm{ref}}$ over time, thereby deterministically reducing the probability of the global gate being active. The global annealing schedule or geometric decay is modeled by a piecewise-linear fitting function, as illustrated in Fig. \ref{fig5}(b).

Second, if the global gate is enabled ($s_{\mathrm g}=1$), the stochastic module generates a bank of \emph{local} stochastic gates. The city identifier chosen during the previous insertion is recovered by reading the relevant word‐line of the $16\times16$ spin array (line \ref{alg_line: prev_sel} of the algorithm). 
Using this index, the FSM issues a row read of the $64\times16$ coupling sub-array; the sense-amplifier bank outputs the encoded inter-city distances, which are realigned into \(4\)-bit words ($d_i$'s) by the shift-and-add (wire re-router) unit.
For each city~$i$, a separate $4$-bit random word, $r_i$, is compared with the corresponding distance value, $d_i$, to form a local stochastic gate $s_i = [\,r_i < d_i\,]$.
This second layer of stochasticity is data-dependent and implements the linear gating in equation \ref{eq:linear_gate} and line \ref{alg_line: local_gate} of the algorithm. 

Random bits for the scheme are generated by an STT‑MRAM‑based true random number generator (TRNG); its circuit schematic is shown in Fig.~\ref{fig5}(c). The core of this design is an MTJ-based differential sense amplifier \cite{mtjsa} that generates a binary output (`0' or `1') by comparing the resistance of its two branches. One branch contains an STT-MTJ whose resistance is actively switched between its parallel state ($R_P$, representing a `1') and its anti-parallel state ($R_{AP}$, representing a `0'). The opposing branch provides a stable reference resistance, $R_\text{ref}$, set to the midpoint value of $(R_{AP}+R_P)/2$. A bi-directional write driver, adapted from \cite{rhs-trng}, maintains the MTJ's switching probability at 50\% and, as a key advantage, eliminates the need for an extra write-cycle for reset. Owing to area constraints, in our design, we consider 16 TRNG units in Fig. \ref{fig5}(c) for generating the bits of $r_g$ in parallel, and a 16 TRNG units for serially generating each bit of $r_i\,\, \forall\,\,i \in \{1, 2, \cdots 16\}$.\\ 

Conventional MTJ‐based random‑bit generators require an additional deterministic reset cycle: they first invoke a stochastic transition \(R_{\text{AP}}\!\rightarrow R_{\text{P}}\) to extract a random bit and then force a deterministic return \(R_{\text{P}}\!\rightarrow R_{\text{AP}}\) to re‑initialize the device for the next random sample.  
In contrast, the proposed driver makes the reverse transition \(R_{\text{P}}\!\rightarrow R_{\text{AP}}\) \emph{stochastic as well}; consequently the sequence $
R_{\text{AP}}\rightarrow R_{\text{P}}\rightarrow R_{\text{AP}}$
yields two independent random bits without any intervening reset phase, thereby halving both write latency and energy per generated bit (see Supplementary Material for detailed timing analysis). To ensure the generation of a true random bitstream that is robust to process variations, we adopt the approach in \cite{rhs-trng} where the final output is produced by XORing the results from two identical units (Fig. \ref{fig5}(d)), which effectively cancels any residual bias from device-to-device variations \cite{rhs-trng}. Fig. \ref{fig5}(e) shows the characteristics of the STT-MTJ used in this work, at its resistance and switching corners. Further details regarding the compact model employed and the fabricated device to which it was calibrated are presented in the Methods section.  \\
While the underlying switching mechanism is an inherently analog process, this approach is significantly more robust than purely analog stochasticity sources. It is easier to control for process and device-to-device variations and is substantially less susceptible to circuit-level analog noise. \\

Finally, the gating unit performs a bit-wise \textsc{and} between \(s_i\) and the mask of currently admissible insertion positions (also maintained by the control block), thereby filtering the candidate list.  
A binary comparator tree selects the city-index with the lowest coupling-matrix value among the surviving candidates (line \ref{alg_line: selection_1} or line \ref{alg_line: selection_2} of the algorithm depending on the value of $s_g$); the winning index is written back to the spin store (line \ref{alg_line: insertion} of the algorithm), completing the iteration for that sub-problem. This selection and stochastic update procedure is then repeated for all remaining sub-problems. Critically, the same global stochastic bit and local 4-bit random words are reused for each sub-problem within this insertion position. At the conclusion of each pass, defined as a full cycle of insertion iterations, the controller evaluates the newly generated tour for each sub-problem and decays $r_{\mathrm{ref}}$. It compares the path length of a given tour against that of the best solution found so far. If the new tour represents an improvement, its configuration is stored in the scratch-SRAM array, replacing the previous best. This entire evaluation and best solution update process is managed asynchronously by the controller. A detailed description of the macro's internal timing and the operation of the FSM-based controller is provided in Supplementary Section 1. 

\subsection{VMM Mode}
\label{subsubsec: vmm_mode}
Analog CiM architectures execute the multiply–accumulate (MAC) operation by summing the pairwise-products of the elements of an input vector and a column of stored weights directly in the analog charge or current domain \cite{shafiee2016isaac, chi2016prime}. Although this in-situ accumulation renders the MAC itself highly energy- and latency-efficient, it necessitates an analog-to-digital converter (ADC) to digitize the accumulated signal; the converter constitutes a dominant share of the peripheral area and power budget \cite{shafiee2016isaac,nag2018newton}.
To curb this overhead, state-of-the-art designs typically provision only a single high-precision ADC (or a small pool thereof) per crossbar array and time-multiplex it across all columns, amortizing silicon cost at the expense of increased read-out latency \cite{wan2022neurram,shafiee2016isaac}. To eliminate the ADC bottleneck, our design foregoes dedicated ADCs and directly quantizes the analog accumulation to a single bit via the sense amplifier array. An input vector is encoded as word-line voltages (RWL and RWLB in Fig. \ref{fig4}(a)), and the accumulation charges the line-capacitance of the bit-line, which is then quantized according to the sign function by the sense-amplifier array. For this scheme to be effective as a neural network primitive, it must support multiplication and signed analog accumulation with ternary weights $\{-1, 0, 1\}$, which is critical for maintaining training stability \cite{psq1, psq2, hcim2}. To implement this capability, we use a push-pull circuit design, illustrated at the bit-cell level in Figure~\ref{fig4}(a). Here, two adjacent bit-cells represent a single ternary weight. The read paths are constructed with an alternating structure: even columns utilize PMOS pull-up transistors, while odd columns use NMOS pull-down transistors. This configuration enables a bi-directional current flow (and charge source) on the source line, directly realizing multiplication with signed weights in the analog domain. During VMMs, the source lines of bit-cells representing the same weight are switched to connect. This design is still compatible with standard SRAM read operations. 
The resulting loss in numerical precision of the VMM analog accumulation is mitigated through hardware-aware training, which maintains classification accuracy on image-classification tasks at parity with software baselines \cite{psq1, psq2}. The training algorithm is described in section 
\ref{subsec: adcless_algo}.

\subsection{Divide and Conquer Algorithm for Large-Scale TSPs}
\label{subsec: dnc_large_tsp}
A common strategy for solving large-scale TSPs is the divide-and-conquer paradigm \cite{yoo2025taxitravelingsalesmanproblem, Lu2024DigitalCIM, hvc}. This approach typically involves decomposing the problem into smaller sub-problems, each small enough to be handled efficiently by a dedicated sub-solver (the LIMO macro in our case), solving them in parallel, and subsequently merging the partial solutions to form a complete tour.

\begin{figure}[h]
\centering
\includegraphics[width=0.71\textwidth]{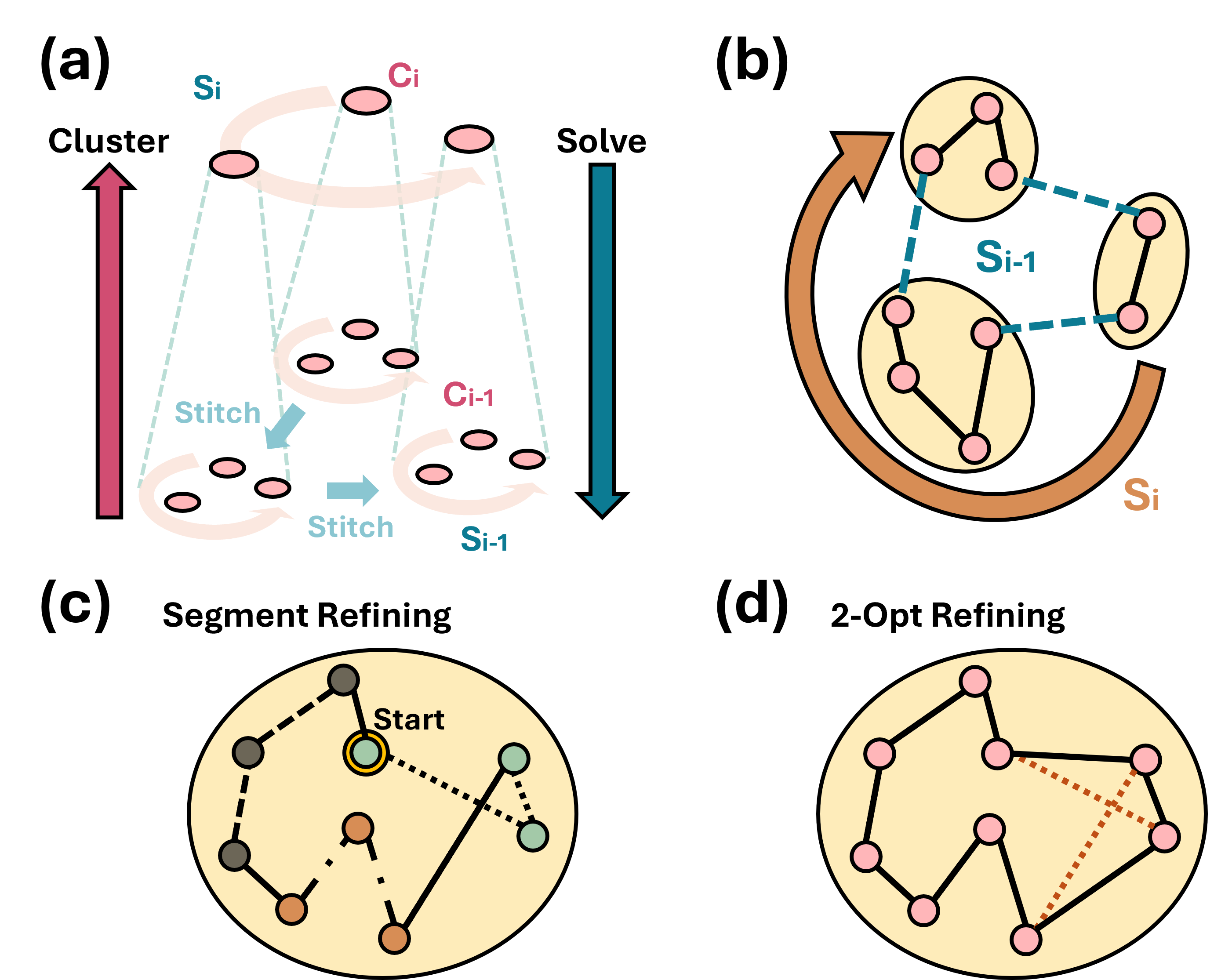}
\caption{Various steps involved in our divide-and-conquer framework for solving large scale TSPs. (a) Bottom-up clustering process and top-down solving process for two hierarchical levels. (b) Tour stitching process for the meta-centroids/nodes at level $i-1$, based on the TSP ordering of centroids in level $i$. (c) Parallelized segment refinement process at level $i$. (d) 2-opt local refinement process at level $i$.}\label{fig7}
\end{figure}

A hierarchical clustering strategy is usually employed to decompose the TSP. The procedure begins by partitioning all nodes of the tour into a set of spatially coherent clusters, often using distance-based metrics such as K-means clustering. The centroid of each cluster is then treated as a TSP meta-node. These meta-nodes are themselves clustered to form higher-level clusters, and this process is applied recursively until the number of meta-centroids in the final level is reduced to a size that can be satisfactorily solved by a chosen sub-solver. After the hierarchical clustering, the algorithm proceeds to solve TSPs of meta-centroids in a top-down manner: the tour over the meta-centroids establishes a coarse global ordering or direction, which is used as a constraint when solving each successive lower level of clusters (Fig. \ref{fig7}(a)). At each hierarchy level the algorithm first determines, for every pair of clusters that are adjacent in the higher‑level tour, the boundary cities with the smallest inter‑cluster distance; those two cities become the designated entry and exit points of their respective clusters. Once these links are fixed, all clusters are dispatched to the sub‑solver in parallel: each sub‑solver computes an open‑loop TSP path that starts at the entry city and ends at the exit city of that cluster (Fig. \ref{fig7}(b)). The resulting sub‑tours are finally concatenated in the sequence dictated by the upper‑level tour, preserving the established global direction. This process of stitching and parallelized sub-tour solving continues down to the bottom level, where the original nodes are arranged into a complete tour. We apply the same hierarchical decomposition strategy in our algorithm, which renders most sub‑tour TSP solutions mutually independent and thus readily parallelizable across multiple LIMO macros. \\

HVC \cite{hvc}, the original paper that proposed the above hierarchical recursive clustering framework for TSPs, employs K‑means clustering at every stage of its hierarchy. However, K‑means presupposes a user‑specified cluster count, necessitating an additional search to determine the appropriate value of $k$ for each level. Moreover, the resulting clusters tend to have lower cohesion, which in turn degrades overall tour quality. TAXI \cite{yoo2025taxitravelingsalesmanproblem} replaces K‑means with agglomerative clustering, yielding tighter clusters and improved solutions; however, this choice substantially lengthens execution time and still demands trial‑and‑error tuning to ascertain the optimal number of clusters at each hierarchical level.  

\begin{algorithm}[t]
\caption{Hierarchical–Clustering based TSP Solver with Refinement}
\label{alg:HierTSP}
\begin{algorithmic}[1]
\Procedure{HierarchicalTSP}{$P,\;T,\;N_{\text{refine}}$}
    \Comment{$P$: coordinates, $T$: cluster‐size threshold, $N_{\text{refine}}$: Number of refinement passes}
    \State $\{C^{0},C^{1},\dots,C^{L}\}\gets\Call{BuildHierarchy}{P,T}$
      \Comment{Top level $C^{L}$ has $|C^{L}|<T$ clusters}
    \State $\pi^{L}\gets\Call{SolveTSP}{C^{L}}$
      \Comment{Initial tour over top‐level meta-centroids}
    \State $\pi^{L}\gets\Call{RefineSegments}{\pi^{L},M_{\text{seg}},P_{\text{passes}}}$

    \For{$\ell=L-1$ \textbf{downto} $0$} \Comment{Descend through hierarchy}
        \State $\{\text{entry},\text{exit}\}\gets
               \Call{FixLinks}{\pi^{\ell+1},C^{\ell}}$
               \Comment{Bind entry/exit cities for each cluster}
        \ForAll{\textbf{in parallel} $C\in C^{\ell}$} \Comment{Parallelized open-TSP solving for each cluster}
            \State $\tau(C)\gets\Call{SolveOpenTSP}{C,\text{entry}(C),\text{exit}(C)}$ \Comment{Solved by LIMO macro}
            \State $\tau(C)\gets\Call{RefineSegments}{\tau(C),P_{\text{passes}}}$
            \State $\tau(C)\gets\Call{TwoOpt}{\tau(C)}$
        \EndFor
        \State $\pi^{\ell}\gets\text{Concatenate }\tau(C)\text{ in the order of }\pi^{\ell+1}$
          \Comment{Stitch sub‐tours to form next‐level tour}
    \EndFor
    \State \Return $\pi^{0}$ \Comment{Closed tour over all original points}
\EndProcedure
\vspace{0.6em}
\Function{BuildHierarchy}{$P,T$}
    \State $C^{0}\gets$ singletons of $P$;\; $\ell\gets0$
    \While{$|C^{\ell}|\ge T$}
        \State $C^{\ell+1}\gets\textsc{Cluster}(C^{\ell})$
          \Comment{PCA based clustering}
        \State $\ell\gets\ell+1$
    \EndWhile
    \State \Return $\{C^{0},\dots,C^{\ell}\}$
\EndFunction
\vspace{0.6em}
\Function{FixLinks}{$\pi,\;C$}
    \ForAll{adjacent clusters $(A,B)$ in $\pi$}
        \State $(a^{*},b^{*})\gets$ closest inter‐cluster pair
        \State $\text{exit}(A)\gets a^{*}$;\; $\text{entry}(B)\gets b^{*}$
          \Comment{Entry/exit pins for a cluster}
    \EndFor
    \State \Return $\{\text{entry},\text{exit}\}$
\EndFunction
\end{algorithmic}
\end{algorithm}

To address the tuning overhead and runtime penalties of earlier hierarchical schemes, we present an enhanced divide‑and‑conquer algorithm, summarized in algorithm \ref{alg:HierTSP}. Its foundation remains the parallel, top‑down cluster‑and‑stitch procedure outlined above, but three key modifications markedly improve solution quality and time-to-solution.

First, cluster formation is performed by a lightweight principal‑component (PCA) bisection routine that recursively partitions the node set until every cluster contains fewer than $T=16$ cities, the largest TSP size solvable by a single LIMO macro. The algorithm is empirically faster than K‑means and more importantly eliminates the need to search for an optimal cluster count at each level (see Supplementary Section 3 for more details).

Second, two refinement phases are executed at every hierarchy level to compensate for the reduced geometric cohesion of PCA partitions. After the entry and exit vertices for each pair of sibling clusters have been designated, the algorithm initiates a parallel segment‑refinement stage. The global tour is partitioned into contiguous subsequences of fixed length $T$, beginning from randomly selected offset nodes, and each subsequence is re‑optimized independently by a LIMO macro running algorithm \ref{alg:SWAI}. Figure \ref{fig7}(c) illustrates this process, where nodes belonging to different subsequences are distinguished by color.  Because the optimization tasks are mutually independent, they can be distributed across multiple macros, rendering the overall computational cost negligible when a sufficient number of LIMO macros are available.

We then apply a deterministic CPU-based 2-opt local‑search pass \cite{croes1958method} to remove residual edge crossings and further shorten the tour at each level (Fig. \ref{fig7}(d). 2-opt repeatedly selects two non‑consecutive edges, \((a,b)\) and \((c,d)\), and replaces them with \((a,c)\) and \((b,d)\); if the substitution reduces the total path length, the intermediate segment is reversed and the change is accepted.  
This process is iterated until no further improving exchanges exist, producing a tour that is locally optimal with respect to all pairwise edge swaps. To limit computational cost, we adopt a standard \(K\)‑nearest‑neighbour variant: for each city, the algorithm considers swaps only with its \(K\) closest neighbours, where \(K \le 20\).  
This restriction lowers the per‑sweep complexity to \(\mathcal{O}(nK)\) from \(\mathcal{O}(n^2)\) while preserving almost all of the quality gains of full 2-opt passes, because the tours are already partially refined.

\subsection{Neural Network Inference Leveraging LIMO Macros}
\label{subsec: adcless_algo}
The VMMs supported by the LIMO macro can also be utilized as a primitive for neural network inference, provided that a hardware-aware training scheme is employed to compensate for the 1-bit partial sum quantization. For this purpose, we implement the methodology from Saxena \emph{et al.} \cite{psq1}, which consists of the following key steps:
\begin{enumerate}
    \item \textbf{Weight and Activation Quantization:} Learned Step Quantization \cite{EsserMBAM20} to map real-valued tensor elements \(\mathbf{x}\) onto fixed-point integers. Given a learned scale factor \(\alpha\), each value is first normalized and clipped to an interval \([Q_{\mathrm{N}},Q_{\mathrm{P}}]\), then rounded down to the nearest integer:
\[
  \mathbf{x}_{\mathrm{int}}
  = \bigl\lfloor
      \mathrm{clip}\!\bigl(\mathbf{x} / \alpha,\; Q_{\mathrm{N}},\,Q_{\mathrm{P}}\bigr)
    \bigr\rfloor, \qquad \mathbf{x}_{\text{quant}} = \mathbf{x}_{\text{int}}\cdot\alpha
\]
Let \(B\) denote the required bit-width. The clipping bounds are then chosen according to the data type:
\[
  \begin{cases}
    Q_{\mathrm{N}} = 0,\quad Q_{\mathrm{P}} = 2^{B}-1,
      & \text{for unsigned activations},\\[6pt]
    Q_{\mathrm{N}} = -2^{B-1},\quad Q_{\mathrm{P}} = 2^{B-1}-1,
      & \text{for signed weights}.
  \end{cases}
\]
\item \textbf{Bit-Slicing:} Weight tensors \(\mathbf{W}\) are decomposed into bit‑planes and mapped onto the crossbar arrays, whereas the activation tensor \(\mathbf{x}\) is streamed in bit‑serially. Since the bit‑slicing operation is non‑differentiable, its gradient is approximated with the straight‑through estimator (STE):
\[
\mathbf{x}_{b} = \operatorname{BitSlice}(\mathbf{x}_{\mathrm{int}}),
\qquad
\nabla_{\mathbf{x}_{\mathrm{int}}}
= \frac{1}{n_b}\sum_{i=0}^{n_b-1}
\frac{\nabla_{\mathbf{x}_{b_i}}}{\bigl(2^{\,s_x}\bigr)^{i}} 
\]
\item \textbf{Mapping Convolutional Layers onto Crossbars:} To deploy a convolutional layer on an Xbar‐based CiM array, the four‑dimensional weight tensor  
\(\mathbf{W}\in\mathbb{R}^{C_{\mathrm{out}}\times C_{\mathrm{in}}\times K_{h}\times K_{w}}\)  
is first reshaped into a two‑dimensional weight matrix  
\(\mathbf{W}_{\mathrm{mat}}\in\mathbb{R}^{C_{\mathrm{out}}\times (C_{\mathrm{in}}K_{h}K_{w})}\)  
by concatenating the spatial and input‑channel dimensions.  An \texttt{im2col} transformation \cite{psq1} likewise flattens each sliding activation window  
\(\mathbf{A}\in\mathbb{R}^{C_{\mathrm{in}}\times K_{h}\times K_{w}}\)  
into an \emph{activation vector}  
\(\mathbf{a}_{\mathrm{vec}}\in\mathbb{R}^{C_{\mathrm{in}}K_{h}K_{w}}\).  With these two steps the convolution reduces to a sequence of VMMs:  
\[
\mathbf{o} \;=\; \mathbf{W}_{\mathrm{mat}}\;\mathbf{a}_{\mathrm{vec}} .
\]

Since \(\mathbf{W}_{\mathrm{mat}}\) and \(\mathbf{a}_{\mathrm{vec}}\) are typically larger than a single crossbar, both operands are tiled to fit the Xbar dimensions. Each tile is bit-sliced as outlined in the preceding section.

Partial sums produced across bit planes, column groups, and time steps are multiplied by their relative significance weights and accumulated spatially (across parallel crossbars) and temporally (over successive input bit-streams). The final accumulation reconstructs the full‑precision output feature map.

\item \textbf{Partial-Sum Quantization}: The crossbar array computes VMMs at 1‑bit partial‑sum quantization, retaining only the sign of the analog accumulation. A learnable scale factor \(\beta_i\) is applied per column to restore the output’s dynamic range. Specifically, for each column \(i\):
\[
p_{q,i}
= \beta_i \,\mathrm{sign}\bigl(p_{\mathrm{analog},i}\bigr)
= \begin{cases}
+\beta_i, & p_{\mathrm{analog},i} > 0,\\
-\beta_i, & p_{\mathrm{analog},i} < 0,
\end{cases}
\]
where \(p_{\mathrm{analog},i}\) denotes the analog‑domain accumulation for column \(i\). During the backward pass, the gradients with respect to the analog partial‑sum \(p_{\mathrm{analog}}\) and the scale factor \(\beta\) are computed as follows:
\begin{equation}
  \nabla_{p_{\mathrm{analog}}}
  = \nabla_{p_q}
  \times
  \begin{cases}
    1, & -1 \;\le\; \dfrac{p_{\mathrm{analog}}}{\beta} \;\le\; 1,\\[6pt]
    0, & \text{otherwise},
  \end{cases}
\end{equation}
\begin{equation}
  \nabla_{\alpha}
  = \nabla_{p_q}
  \times
  \operatorname{sign}\!\Bigl(\dfrac{p_{\mathrm{analog}}}{\beta}\Bigr),
\end{equation}
where \(\nabla_{p_q}\) denotes the upstream gradient flowing into the quantized partial‑sum.

\subsection{Circuit Simulations}
\label{subsec: circuit sims}
Our circuit evaluation methodology encompassed both macro-level performance analysis and block-level physical validation. The LIMO macro was designed from the transistor level using a 65nm process and simulated at 100MHz to determine its power consumption and latency for a annealed insertions and VMM operations. Concurrently, we generated the physical layouts for all constituent digital and analog blocks for macro area estimations. The analog components in particular, namely the STT-TRNG and the sense-amplifier array, were subjected to post-layout testing in isolation to confirm their correct operation under parasitics and MTJ process corners (for the TRNG). Supplementary Section 2 shows the layouts for the analog peripherals, along with an area breakdown of the macro components.

\begin{figure}[h]
\centering
\includegraphics[width=0.8\textwidth]{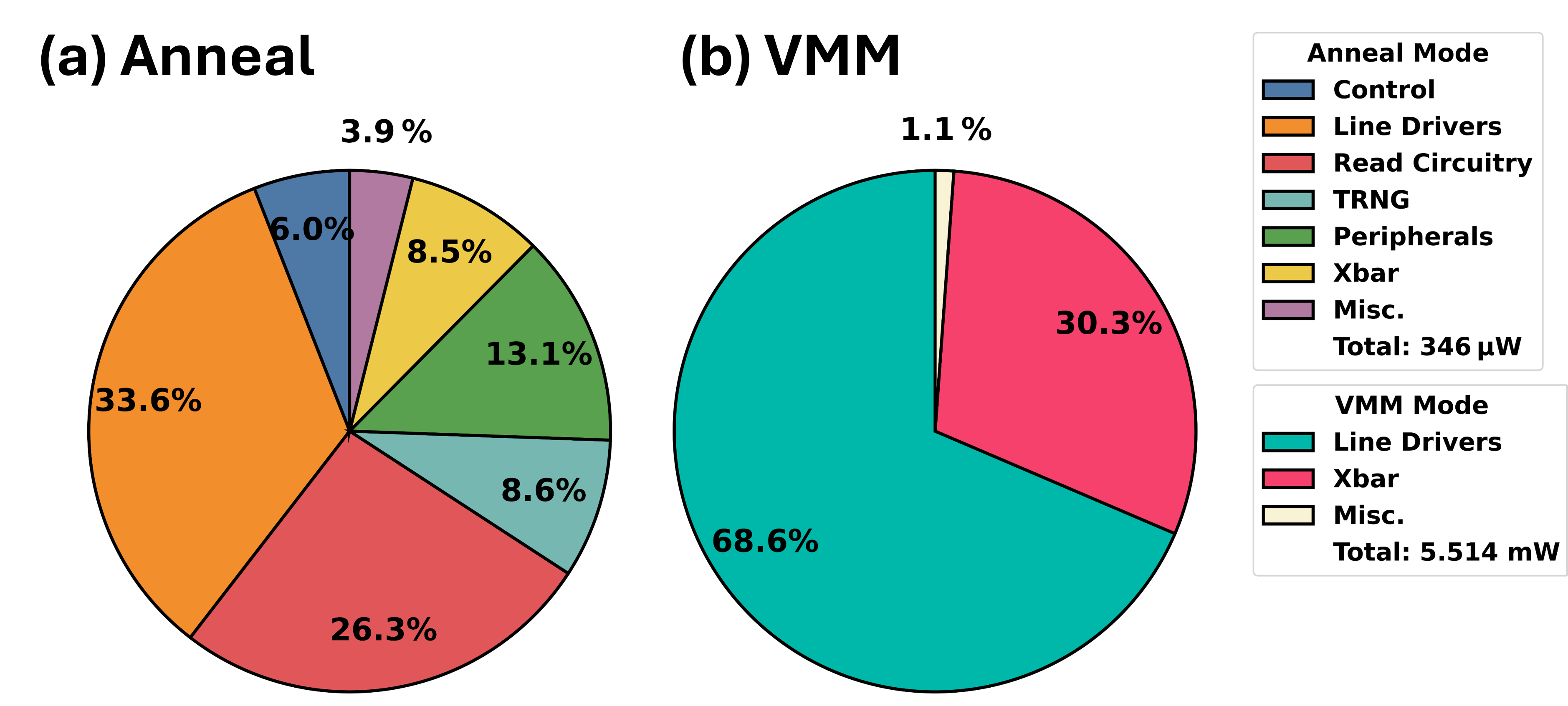}
\caption{Breakdown of power consumption among the LIMO macro's components when operating in (a) annealing mode and (b) VMM mode at 100MHz.}\label{fig6}
\end{figure}
\begin{table}[h]
  \caption{Comparison of various annealers in literature}\label{tab:ising_bench}
  \begin{tabular*}{\textwidth}{@{\extracolsep\fill} l l l r l r r}
    \toprule
    Annealer   & Implementation            & Task    & Spins & Connectivity    & Scaled Power & Scaled Area \\
               &                           &         &       &                 & (mW)\footnotemark[1] & (mm$^2$)\footnotemark[1] \\
    \midrule
    STATICA\cite{STATICA}    & Taped‐out, 65\,nm         & Max‐Cut & 512   & Fully connected & 1.267        & 0.02344      \\
    Ping‐Pong\cite{pingpong}  & Taped‐out, 65\,nm         & Max‐Cut & 100   & Fully connected & 0.042        & 0.0197       \\
    SMTJ‐A\cite{Si2024EnergyEfficient}     & Taped‐out, 40\,nm/SMTJ    & TSP & 80    & Fully connected & 0.008        & 10           \\
    DCIM‐A\cite{Lu2024DigitalCIM}     & Simulated, 16/14\,nm      & TSP     & --    & Clustered       & 604          & 22.085       \\
    TAXI\cite{yoo2025taxitravelingsalesmanproblem}       & Simulated, 65\,nm/SOT‐MTJ & TSP     & 144   & Fully connected & 0.036        & --           \\
    \textbf{LIMO}       & Simulated, 65\,nm/STT‐MTJ & TSP     & 1280  &  Fully connected \footnotemark[2] & 0.00135      & 1.7E-04 \footnotemark[3]          \\
    \botrule
  \end{tabular*}
  \footnotetext[1]{Scaled area (area per physical spin) and scaled power (power per physical spin) are computed as the actual quantity divided by (number of spins $\times$ connectivity ratio). Connectivity ratio is defined as the number of physical connections divided by the number of connections in an equivalent fully connected graph.}
  \footnotetext[2]{5 independent 16$\times$16 fully connected graphs, implying a connectivity ratio of 0.2. }
  \footnotetext[3]{Actual area of a LIMO macro is $0.044\text{mm}^2$}
\end{table}

Fig. \ref{fig6} shows the power distribution of various components and the total power consumed by a LIMO macro while executing the two operating modes. The macro's latency is mode-dependent, requiring an average of 25.4 clock cycles for a single insertion step in annealing mode (processing five sub-problems concurrently), and one clock cycle for a partial-sum quantized VMM. Table~\ref{tab:ising_bench} presents a comparison of the LIMO macro with several hardware annealers from the literature, highlighting LIMO's superior efficiency in terms of its power-per-effective-spin and area-per-effective-spin ratios. Our design achieves significant energy efficiency gains over other works by eliminating digital-to-analog converters and utilizing an STT-MTJ-based TRNG module over CMOS-based stochasticity sources. This approach is also area-efficient, as the STT-MTJ module is compact (occupies only 8.5\% of macro area, see supplementary), and the annealing logic is implemented as add-on peripherals to a nearly foundry-standard 8T-SRAM core.

\subsection{System Level Simulations for Large Scale TSPs}
\label{subsec: sys_sim_tsp}
Our proposed algorithm was evaluated on the TSPLib benchmark suite \cite{reinelt1991tsplib}, on instances with up to 85900 nodes, the largest problem available in the repository. Solution quality was compared with state-of-the-art annealers for large-TSPs: NeuroIsing \cite{neuroising}, TAXI \cite{yoo2025taxitravelingsalesmanproblem}, and DCIM-A \cite{Lu2024DigitalCIM}. For large-scale simulations we employed a Python-based model of the LIMO macro that replicates the functional behavior assumed by Algorithm \ref{alg:SWAI} (see Methods for details). 
\begin{figure}[h]
\centering
\includegraphics[width=\textwidth]{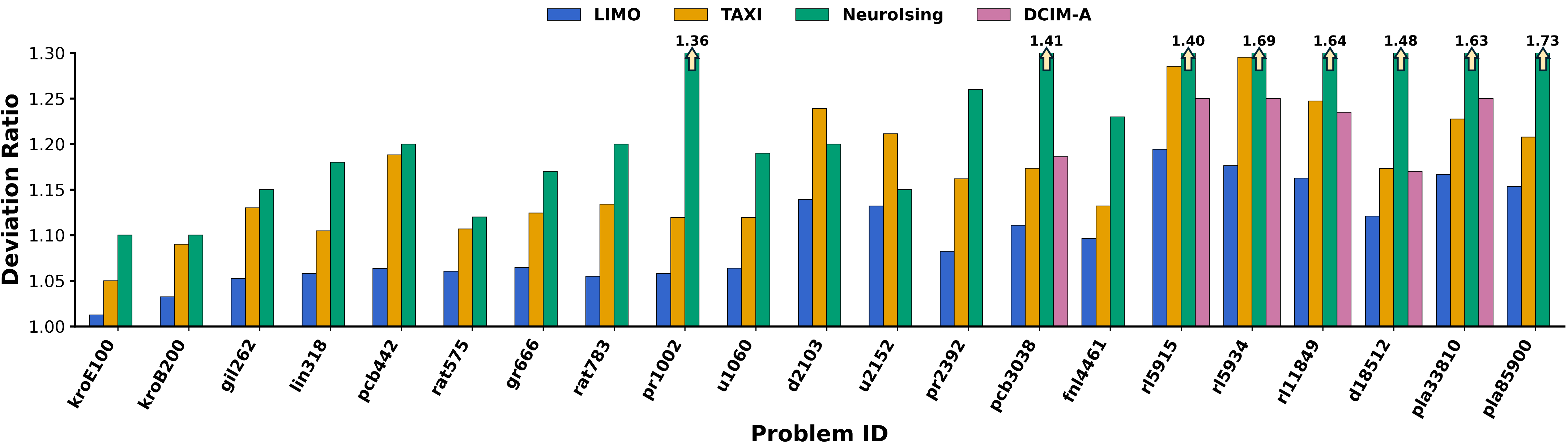}
\caption{A comparison of deviation ratios for various TSP-annealers in literature. The deviation is measured relative to optimal solutions from exact solvers, with lower values indicating superior performance. The numerical value in each problem ID denotes the number of nodes of the TSP.}\label{fig8}
\end{figure}
\\ 
As demonstrated in Figure~\ref{fig8}, our algorithm leveraging LIMO macros consistently outperforms other annealers from the literature, achieving state-of-the-art solution quality across all benchmarked problems. The average improvement over TAXI in the deviation ratio across all considered problems is 37.5\%. The two refinement steps contribute substantially to the final solution quality, compensating for the relatively sub-optimal PCA-clustering. The annealing hyperparameters for each problem instance can be found in the Methods section. Ablations of each refinement step for each problem ID in Fig. \ref{fig8} is presented in Supplementary Section 4.
\begin{figure}[h]
\centering
\includegraphics[width=0.9\textwidth]{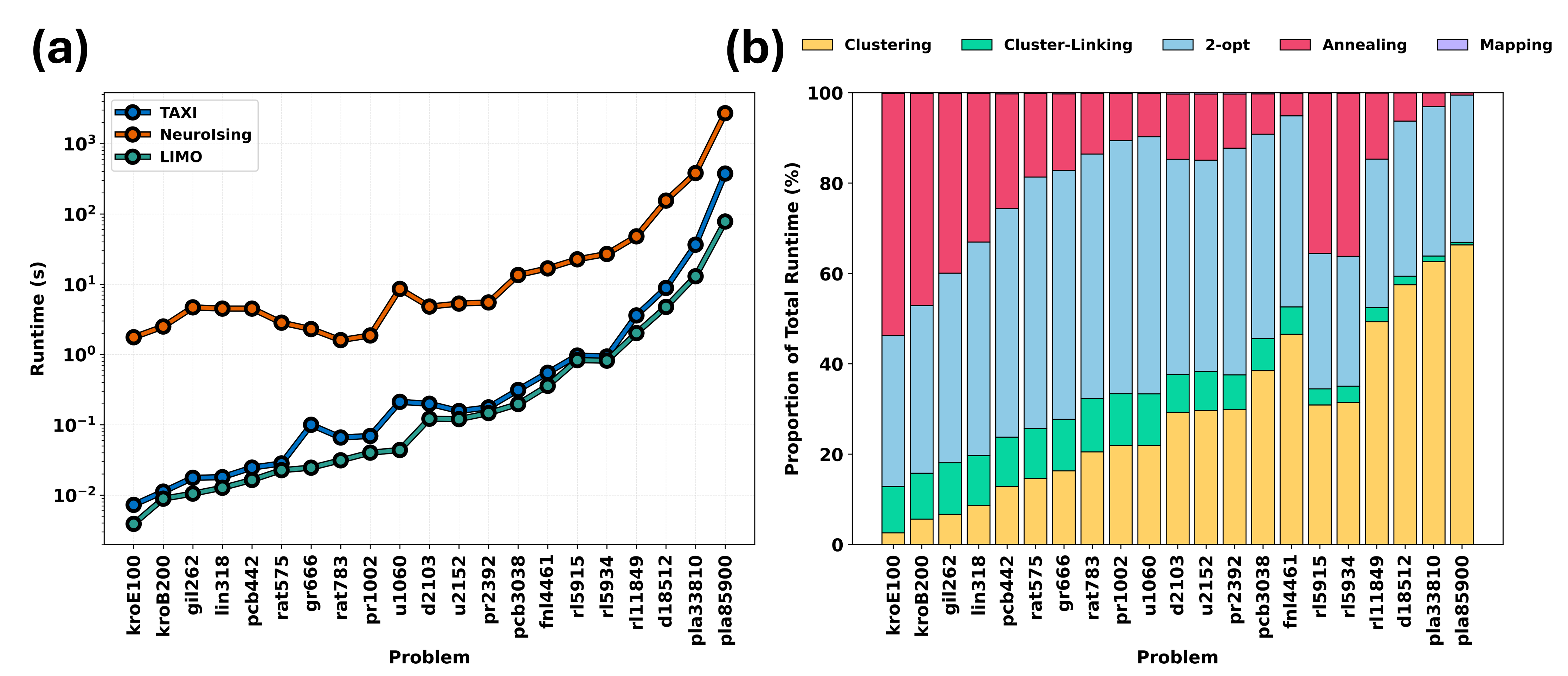}
\caption{(a) TSP Solver runtime comparison: LIMO vs. TAXI vs. NeuroIsing. (b) Runtime profiling of the LIMO algorithm, showing the percentage contribution of each constituent step for the same problem instances.}\label{fig9}
\end{figure}
\\
To evaluate the system-level performance of our algorithm, we simulated a spatial accelerator architecture composed of multiple LIMO macros. This accounts for practical architecture-level overheads, including the latency associated with mapping data from off-chip memory onto the macros. The simulation framework is described further in the Methods section.

Fig. \ref{fig9}(a) shows a total runtime comparison of annealer based TSP solvers on instances of varying size. Most notably, for the largest instance, a TSP with 85900 cities, LIMO is $\sim5\times$ faster than the former state-of-the-art, while delivering superior solution quality on the same instance. Clustering is a primary computational bottleneck for TAXI and NeuroIsing. For instance, it accounts for 99\% of TAXI's total runtime on the 85,900-city instance. Our lightweight PCA-based clustering routine addresses this bottleneck, offering improved runtimes as a result. It is also important to note that the timings for TAXI do not include the search process for optimal number of clusters at each hierarchy, which usually requires multiple runs. Fig. \ref{fig9}(b) shows the runtime distribution of the various steps in our algorithm. 
\subsection{System Level Simulations for CNN Workloads}
\label{subsec: sys_sim_cnn}
With the forward and backward pass operations defined for our hardware in section \ref{subsec: adcless_algo}, we proceed to evaluate its performance on practical CNN inference. To this end, we train two variants of the Resnet20 architecture \cite{HeZRS16}: one for image classification on the CIFAR-10 dataset \cite{krizhevsky2009learning} and another for face detection on the PASCAL dataset (called ResnetSSD) \cite{EveringhamGWWZ10}. The system-level latency and energy consumption for these inference tasks are then assessed using a cycle-accurate spatial architecture simulator. For a comparative analysis, we establish a baseline using a conventional SAR-ADC-based crossbar-array based spatial architecture of equivalent size to our LIMO macro, that is optimized for energy efficiency \cite{7063128}. A detailed description of the datasets, network models, training procedure, and baseline architecture parameters is provided in the Methods section. Table~\ref{tab:adcless_acc} presents the inference accuracy on the CIFAR-10 and PASCAL datasets, demonstrating performance that approaches the software baseline due to the use of CiM-aware training. 

\begin{table}[h]
\caption{Inference accuracy for considered edge datasets compared with baselines}\label{tab:adcless_acc}
\begin{tabular*}{\textwidth}{@{\extracolsep\fill}cccccc}
\toprule
\multicolumn{3}{@{}c@{}}{\textbf{Resnet20 -- CIFAR10}} &
\multicolumn{3}{@{}c@{}}{\textbf{ResnetSSD -- Face Detection}} \\\cmidrule(lr){1-3}\cmidrule(lr){4-6}
\shortstack{\textbf{Software}\\\textbf{Baseline}} &
\shortstack{\textbf{Weights/Activations}\\\textbf{Quantized}} &
\shortstack{\textbf{LIMO}\\\textbf{(ADC-Less)}} &
\shortstack{\textbf{Software}\\\textbf{Baseline}} &
\shortstack{\textbf{Weights/Activations}\\\textbf{Quantized}} &
\shortstack{\textbf{LIMO}\\\textbf{(ADC-Less)}} \\
\midrule
89.6\% & 89.5\% & 89.3\% & 97.8\% & 97.71\% & 95.69\% \\
\botrule
\end{tabular*}
\end{table}

\begin{figure}[h]
\centering
\includegraphics[width=0.9\textwidth]{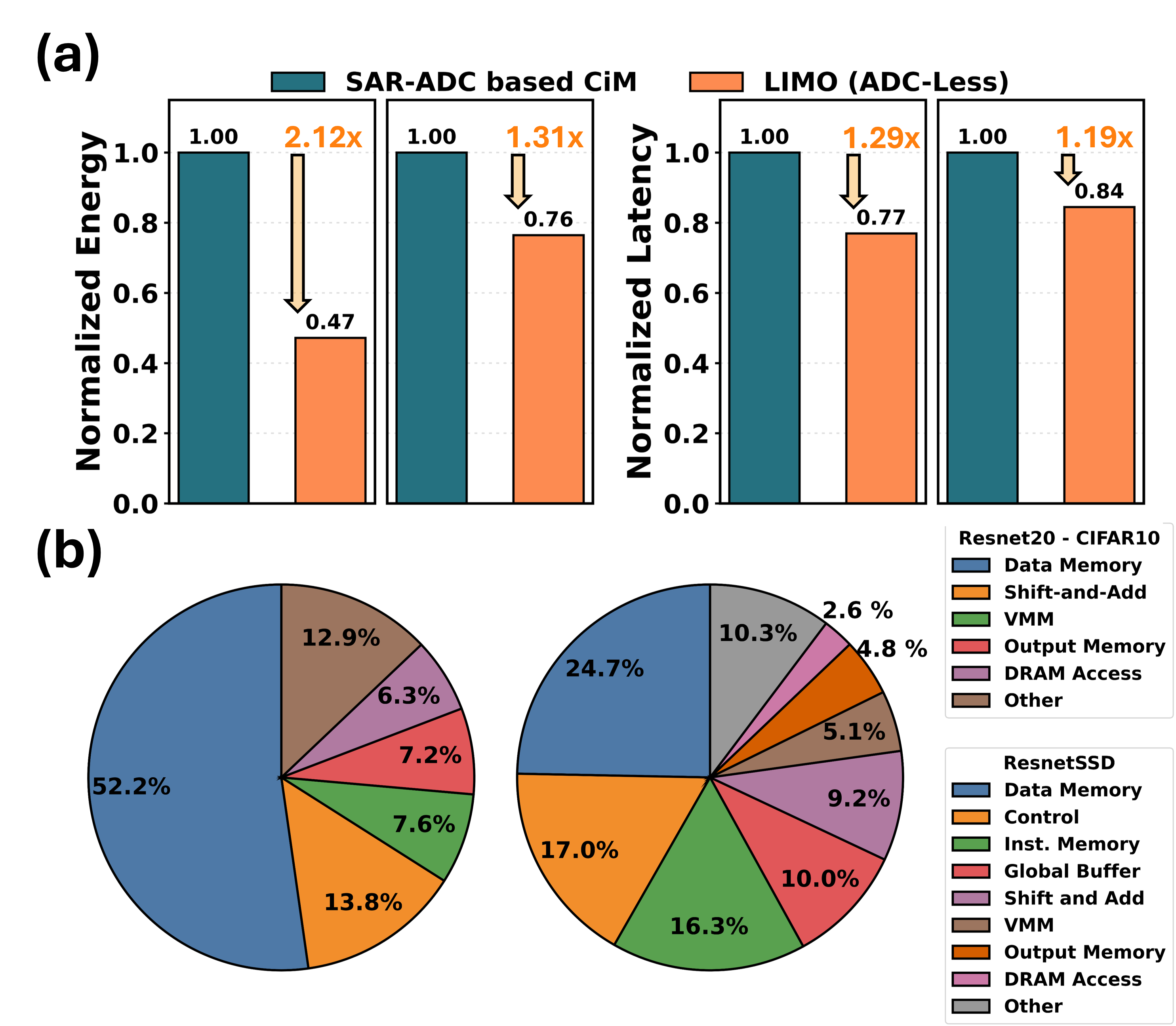}
\caption{(a) Comparison of normalized latency and energy consumption between our effectively ADC-less LIMO macro and a conventional CiM array that utilizes an energy-efficient SAR-ADC. Results are shown for the two types of CNNs considered: Resnet20 (for CIFAR-10 image classification) and ResnetSSD (for face detection). (b) Breakdown of energy consumption by component for the PUMA-based spatial architecture incorporating our macro, for the two CNNs.}\label{fig10}
\end{figure}
Figure~\ref{fig10} details the system-level performance of our ADC-less CiM macro, presenting (a) the corresponding gains in energy and latency efficiency and (b) a breakdown of the architecture-level energy consumption for the two considered workloads. The energy improvement is due to sense-amplifiers consuming less energy than the SAR-ADCs in the baseline. The latency improvement is due to our single-pass analog-to-digital conversion, enabled by dedicated per-column sense amplifiers, which avoids the time-multiplexing required by the SAR-ADC-based CiM.

\end{enumerate}

\section{Discussion}\label{sec12}
In this article, we presented LIMO, a low-power, mixed-signal computational macro designed to accelerate annealing-based combinatorial optimization within a spatial architecture. Its annealing capability is enabled through dedicated in-memory peripherals, particularly an STT-MRAM-based TRNG serving as a compact and energy-efficient source of on-chip stochasticity. The modular nature of the annealing peripherals imply that the core 8T-SRAM array can be reused to perform VMMs, which in turn allows the same architecture to serve NN inference. Owing to pipelined macro-level dataflows, modular annealing peripherals and an energy efficient stochastic unit, our macro exhibits an operating power approximately $15\times$ lower than prior state-of-the-art hardware annealers. 

 The hardware-algorithm co-design of a system leveraging multiple LIMO macros achieves state-of-the-art solution quality on large-scale TSP instances. This stems from our two-pronged algorithmic approach. At the macro level, our significance-weighted annealed insertion scheme yields superior sub-problem solutions. At the system level, this is complemented by a hierarchical clustering strategy that divides a large instance for parallel solving, followed by refinement iterations to optimize the globally constructed tour. A clustering bottleneck prevalent in previous annealers is addressed by a PCA-based bisection scheme, which yields an $\sim5\times$ faster time-to-solution on an 85,900-city TSP than the prior art. \\
Through hardware-aware training to support neural network inference, our system achieves accuracy on image classification and face detection tasks that is comparable to software baselines. Concurrently, this is accomplished with improvements in hardware efficiency over baseline CiM architectures, owing to an ADC-less approach. \\
 Future research can explore deeper architectural and algorithmic synergies, especially in reducing the system's reliance on general-purpose processors for annealing, integrating neural-combinatorial optimization techniques on our platform, and extending our approach to a broader range of workloads limited by the memory wall.

\section{Methods}\label{sec11}
\subsection{STT-MTJ Characterization}
\begin{figure}[h]
\centering
\includegraphics[width=1\textwidth]{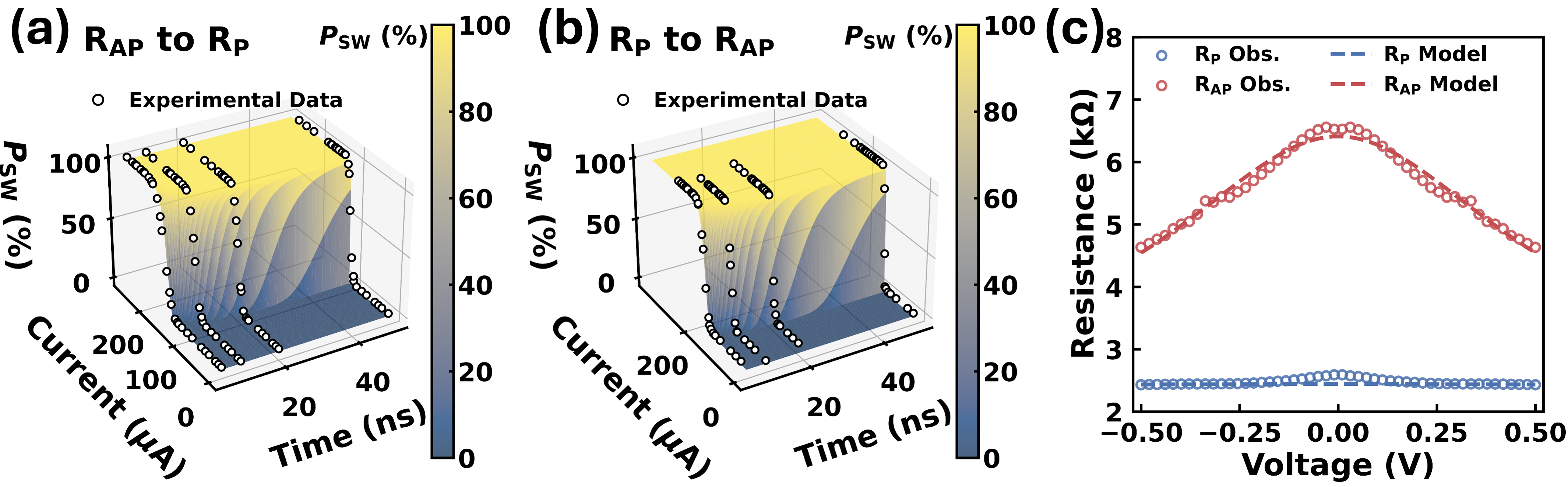}
\caption{Characteristics of the MTJ as predicted by the compact model \cite{vlsisttdevice2025}, which has been calibrated to experimental data. (a, b) Nominal (TT corner) switching probabilities for the $R_{AP} \rightarrow R_{P}$ and $R_{P} \rightarrow R_{AP}$
  transitions, respectively, plotted as a function of write current amplitude and pulse width. (c) Nominal resistance values for the two MTJ states, as determined by the model's calibration to experimental R-V characteristics.}\label{fig11}
\end{figure}

The MTJ considered in this work is a perpendicular STT-MTJ exhibiting a nominal tunnel magnetoresistance (TMR) of 163\%, and a 60nm diameter \cite{imecmtj}. From bottom to top, the stack comprises a buffer/seed layer, a synthetic antiferromagnet (SAF) together with an adjacent reference layer that provides a fixed out-of-plane magnetization; an ultra-thin MgO tunnel barrier, and a composite free layer configured as FL1/Ta/FL2. The stack is terminated with a cap layer. \\
Following the physical characterization of the device, we employ a compact SPICE model to accurately simulate its behavior at the circuit-level. More details of the compact model, and the corners tested on are given in section \ref{subsubsec: sims_hyperparams}. The switching probability characteristics and the Resistance-Voltage (R-V) characteristics of the compact model along with the experimental data it was calibrated on (for the nominal corner), is shown in Fig. \ref{fig11}.

\subsection{Circuit Design and Simulation Framework}
All constituent blocks of the LIMO macro were designed using the TSMC 65nm process in Cadence Virtuoso. Circuit simulations were carried out in the Cadence Spectre simulator. The digital blocks (Control block, decoders, drivers, shift and add unit, gating unit, and the comparator tree) were synthesized from RTL code using Genus Synthesis Solution, and laid out using Innovus Implementation System. Layouts for the analog blocks (8T-SRAM CiM macro, scratch 6T-SRAM, stochastic module, and the sense amplifiers) were made manually in Virtuoso Layout XL. It is assumed that the CMOS circuitry is at the front-end-of-the-line ($<$ Metal 3) with the STT-MTJs bonded on top at the back-end-of-the-line ($>$ Metal 3).
\subsection{Cycle Accurate Spatial Architecture Simulations}
System-level evaluations for both large-scale TSP annealing and CNN inference were carried out using the publicly available cycle-accurate PUMA simulator \cite{puma}. The system is organized as a hierarchical spatial architecture. At the top level, multiple tiles are interconnected by a network-on-chip, while a global buffer facilitates communication with off-chip DRAM. Each tile, in turn, is composed of instruction memory, shared memory, and several processing cores. Within each core, there are multiple compute-units, a register file, and a dedicated memory controller. The compute-unit is composed of several LIMO macros, which replace the ReRAM-based CiM units that were in the original PUMA architecture. A compiler generates instructions and maps the sub-TSPs or tiled weight matrices to the LIMO macros. To ensure compatibility with the simulator, which is based on a 32nm process, our 65nm circuit-level simulation results were scaled using the predictive scaling equations presented in \cite{STILLMAKER201774}.\\ 
\noindent\textbf{1.\;Decompose the measured energy:}
For both VMM and annealing operating modes we first split the total energy recorded in 65nm into its digital and analog components:
\[
E_{65}^{\mathrm{dig}}=\gamma_{\mathrm{dig}}E_{65},\qquad
E_{65}^{\mathrm{ana}}=\gamma_{\mathrm{ana}}E_{65},\qquad
\gamma_{\mathrm{dig}}+\gamma_{\mathrm{ana}}=1.
\]

\noindent\textbf{2.\;Digital‑energy scaling factor:}
Stillmaker \& Baas \cite{STILLMAKER201774} define a quadratic
\emph{EnergyFactor} polynomial for each node,
\begin{equation}
\label{eq:efactor}
\mathrm{EnergyFactor}_{n}(V)=
a^{(n)}_{e2}V^{2}+a^{(n)}_{e1}V+a^{(n)}_{e0}.
\end{equation}
Evaluating this at the common digital supply
$V_{\!{\text{dig}}}=0.8\,$V gives
\[
\eta_{\mathrm{dig}}=
\frac{\mathrm{EnergyFactor}_{32}(V_{\!{\text{dig}}})}
     {\mathrm{EnergyFactor}_{65}(V_{\!{\text{dig}}})},
\qquad
E_{32}^{\mathrm{dig}}=\eta_{\mathrm{dig}}\,E_{65}^{\mathrm{dig}}.
\]

\noindent\textbf{3.\;Analog‑energy scaling factor:}
Assuming capacitive dominance ($E\!\propto\!C\,V^{2}$) and that
capacitance scales with active area, the analog multiplier is
\[
\eta_{\mathrm{ana}}=\alpha_{\mathrm{area}}
\Bigl(\tfrac{V_{\!{\text{ana}},32}}{V_{\!{\text{ana}},65}}\Bigr)^{\!2},
\qquad
E_{32}^{\mathrm{ana}}=\eta_{\mathrm{ana}}\,E_{65}^{\mathrm{ana}},
\]
where $\alpha_{\mathrm{area}}$ is the area‑scaling factor from
Stillmaker \& Baas \cite{STILLMAKER201774} and $V_{\!{\text{ana}}}$ the analog supply voltage (=1.2V for 65nm and 1.1V for 32nm).

\noindent\textbf{4.\;Latency transformation:}
Digital latency scales with clock frequency:
\[
T_{32}^{\mathrm{dig}}=
\frac{T_{65}^{\mathrm{dig}}\,f_{65}}{f_{32}}.
\]
where $f_{65}= 100\text{MHz}$ and $f_{32} = 1\text{GHz}$. The STT-MTJ switching in the stochastic module takes the same amount of time in 65nm as it does in 32nm (10ns), and its latency is unchanged:
$T_{32}^{\mathrm{ana}}=T_{65}^{\mathrm{ana}}$.
Hence
$T_{32}=T_{32}^{\mathrm{dig}}+T_{32}^{\mathrm{ana}}$.

\noindent\textbf{5.\;Aggregate 32nm figures:}
Finally,
\[
E_{32}=E_{32}^{\mathrm{dig}}+E_{32}^{\mathrm{ana}},\qquad
P_{32}=\frac{E_{32}}{T_{32}}.
\]
Because the coefficients of \eqref{eq:efactor} subsume both dynamic and
leakage mechanisms, $E_{32}$ and $P_{32}$
implicitly capture static and switching contributions without further
adjustment. 
\subsection{Divide and Conquer Algorithm for Large-Scale TSPs}
To benchmark our algorithm's solution quality on TSPLib problems, we developed a digital twin of our system in Python. Our framework models both the low-level LIMO macro (for core SWAI algorithm operations) and the high-level divide-and-conquer logic, including its clustering, k-neighborhood two-opt, and cluster linking stages. All experiments for TAXI, NeuroIsing, and LIMO were run on a 32 core AMD\textsuperscript{\textcircled{R}} EPYC 7502 CPU. 

\subsection{Dataset Information for CNN Training and Inference}
The CIFAR‑10 \cite{krizhevsky2009learning} benchmark provides 60000 colored $32\times32$‑pixel images drawn from ten everyday object categories; we train on 50000 images and test on 10000 images. The metric here is classification accuracy on the test set.

For face detection, we train on WIDER FACE \cite{Yang_2016_CVPR}, which contains 32203 images with 393703 annotated faces exhibiting large variations in scale, pose, occlusion and illumination; the dataset is organized into 12880 training, 3226 validation and 16097 test images across 61 event categories, following its official split. In line with papers that train on this dataset, we report inference accuracy on the PASCAL Faces benchmark \cite{EveringhamGWWZ10}, a subset of PASCAL VOC comprising 851 images annotated with 1335 faces. Detection accuracy is reported as Average Precision (AP) computed with the PASCAL VOC‑2007 protocol implemented in the open‑source face‑eval toolkit (\url{https://github.com/sfzhang15/face-eval}). Candidate detections are ranked by confidence, matched to ground‑truth faces at an intersection‑over‑union threshold of 0.5, and the precision–recall curve traced over this ranking is summarized by the 11‑point interpolated area under the curve; because the datasets contain a single class, this AP value coincides with mean AP. 

\subsection{CNN Architecture and Training Framework}
To enable hardware-aware training, we use a custom framework in PyTorch that simulates the execution of convolutional layers on our macros during both the forward and backward passes. This simulation accurately models key hardware-specific behaviors, including weight and activation quantization, weight tiling and activation bit-streaming, partial-sum quantization of our ADCLess crossbars, and bit-plane accumulation across macros. \\
For the two inference tasks, we train two variants of the Resnet20 architecture, one for image classification on CIFAR-10 and the other for face detection on the WIDER FACE dataset (ResnetSSD). Both models are quantized to 3-bit weights and 3-bit activations, except for the first layer of each backbone, which remains in full-precision, following standard practice \cite{psq1, psq2, hcim2}. A single Nvidia H200\textsuperscript{\textcircled{R}} GPU was used to train both networks.\\

\textbf{Image Classification on CIFAR-10}

For this task, we employ the standard Resnet20 architecture for the CIFAR-10 dataset \cite{HeZRS16}.

\textit{Architecture}: The network begins with an initial $3\times3$ convolution, followed by three stages of three residual blocks each. A residual block consists of three convolutional layers with activation functions applied (ReLU in this case). This block employs a ``skip connection" that adds the input of the block to its output, allowing gradients to pass easily. The network concludes with an adaptive average pooling layer and a final fully-connected layer for the 10-class classification.

\vspace{1em} 

\textbf{Face Detection on WIDER FACE}

For this task, we use a detector model composed of a modified Resnet20 backbone and specialized detection heads, designed for a $640 \times 640$ input resolution.

\textit{Architecture}: The same Resnet20 backbone is adapted for face detection, beginning with a 7$\times$7 strided convolution (stride 4) and a max-pooling layer to reduce the high input spatial resolution. This is followed by three main stages, each containing three residual blocks as in the CIFAR-10 variant. To create a feature pyramid for detecting faces at multiple scales, two additional 3$\times$3 strided convolutional layers are appended after the backbone. The final classification and bounding box regression heads are composed of full-precision convolutional layers, with a separate pair of heads applied to each of the three feature maps in the pyramid.

\subsection{Experimental Setup and Parameters}
\label{subsubsec: sims_hyperparams}
This section specifies the parameters for all presented experiments across the stack, from the circuit-level models and operating points to architectural configurations and algorithm hyperparameters. 
\\ \\
\textbf{Circuit Simulations.}

The CMOS components of the 8T-SRAM core, scratch 6T-SRAM, STT-MTJ-based RNG module, and sense amplifiers were simulated using the foundry-provided TSMC 65nm standard power SPICE models. Circuit-level functionality was verified across three process corners (TT, FF, SS) at $V_{DD}=1.2V$ and a 100MHz operating frequency. The FSM control logic, word-line decoders/drivers, and the digital blocks of the annealing peripherals were synthesized from RTL code using the TSMC 65nm Low Power (LP) standard cell library, and used at $V_{DD}=0.8V$ and 100MHz. The digital designs were verified to be free of timing violations at the SS process corner for the aforementioned operating points. 

For the STT-MTJ, we employ a compact SPICE model, based on the 2D Fokker-Planck equations described in \cite{2dfpe}. The model captures the stochastic nature of STT switching by determining the probability of switching as a function of the applied write current amplitude and pulse duration. To ensure accuracy, the model was calibrated against experimental data from the device's RV-curves and switching probability measurements. This calibration was performed across nine process corners to account for device variability. These corners, denoted by an XY convention, describe variations in both switching dynamics (X: Typical, Fast, or Slow) and MTJ resistance (Y: Typical, Low, or High). The device's nominal TMR is 163\%, with nominal parallel ($\text{R}_\text{P}$) and anti-parallel ($\text{R}_\text{AP}$) resistances of $2.44k\Omega$ and $6.41k\Omega$, respectively, measured at $V_{MTJ}=0V$. The RNG-module was tested across all $9\times5$ MTJ+CMOS corners for functional correctness.
\\ \\
\textbf{System Level Parameters.}
Spatial architecture specifications for cycle-accurate simulations as well as the ADC-based CiM baseline specifications are enumerated in table \ref{tab:components}.

\begin{table}[ht]
  \centering
  \caption{Spatial Accelerator (PUMA) Specifications}
  \label{tab:components}
  \begin{tabularx}{\linewidth}{|Y|Y|Y|}
    \hline
    \textbf{Component} & \textbf{Parameter} & \textbf{Value} \\ \hline
    Global Buffer            & Storage      & 256KB \\ \hline
    Shared Memory            & Storage      & 256KB \\ \hline
        Bandwidth                & Paths        & \makecell[l]{%
      \textbf{Tile eDRAM $\rightarrow$ IMA bus} \\ 32~GB/s peak;\;16~GB/s sustained\\
      \textbf{Across tiles (on-chip NoC link)} \\ 4~GB/s peak ,\;$\approx$0.32~GB/s avg/port\\
      \textbf{Off chip (Hyper Transport)} \\6.4~GB/s per node%
    } \\ \hline

    Tile                     & \# of cores  & 8       \\ \hline
    Tile eDRAM       & Parameters       & \makecell[l]{%
      Access: 2~ns\\
      Bus: 256\text{-}bit\\
      Receive buffer: depth 150,\; width 16%
    } \\ \hline

    LIMO Macro               & \# per core  & 16      \\ \hline
    ALU                      & ALU width    & 64      \\ \hline
    Register File (RF)       & Storage      & 4KB    \\ \hline
    VMM Peripherals     & Type &
      \makecell{
      \textbf{ADC-Less}\\
        1-bit, 0.0125ns, 0.0022pJ \\
        \textbf{Energy efficient SAR-ADC}\\
        7-bit, 0.15ns, 0.59pJ \cite{7063128} \footnotemark[1] \\
      } \\ \hline
  \end{tabularx}
  \footnotetext[1]{The reported energy and latency values are on a per-column basis for the crossbar array.}
\end{table}

For benchmarking performance on the TSPLib dataset using our digital twin implemented in Python, we use different hyperparameters depending on the scale of the problem. Until \texttt{u1060}, the annealing hyperparameters are $p_0=0.3,\, \beta=0.995,\, p_{min}=0.05,\, N_{\text{refine}}=10$. Until \texttt{fnl4461}, they are $p_0=0.3,\, \beta=0.995,\, p_{min}=0.05,\, N_{\text{refine}}=30$. For larger problems, they are $p_0=0.2,\, \beta=0.9995,\, p_{min}=0.01,\, N_{\text{refine}}=30$ (all notations follow those in algorithms \ref{alg:SWAI} and \ref{alg:HierTSP}). Setting the maximum number of elements per cluster (fixed at 16 for all instances) automatically determines the total number of clusters and the depth of the hierarchy. For 2-opt based refining, we limit K (the neighborhood pruning parameter) to 20 for all instances.
\\ \\
\textbf{Hardware Aware Neural-Network Training Hyperparameters}
\begin{enumerate}
    \item \textit{Image Recognition (CIFAR-10)}: Resnet20 is trained for 800 epochs, with the first 400 epochs for training the network with quantized weights/activations, and the second stage for training with partial-sum quantization. We use a stochastic gradient descent (SGD) optimizer with Nesterov momentum (0.9), a weight decay of $1 \times 10^{-4}$, and a batch size of 256. The learning rate for both stages begins at 0.01 and follows a cosine annealing schedule. Standard CIFAR-10 data augmentation (random cropping and horizontal flipping) is utilized.
    \item \textit{Face Detection (WIDER-FACE)}: ResnetSSD is trained for 200 epochs, split into a 100-epoch quantized weight/activation training stage and a 100-epoch partial-sum-quantization training stage. We use an SGD optimizer with an initial learning rate of $5 \times 10^{-3}$, momentum of 0.9, and weight decay of $5 \times 10^{-4}$. The learning rate is decayed by a factor of 0.1 at 50\% and 83\% of the epochs within each stage. The training batch size is 32.
\end{enumerate}

\bmhead{Supplementary information}
This article contains supplementary information.

\bmhead{Acknowledgements}
The authors thank Sangmin Yoo, Gaurav Kumar K, and Utkarsh Saxena for technical discussions, and Tanvi Sharma for reviewing the manuscript. This work was supported in part by CHEETA: CMOS+MRAM Hardware for Energy-Efficient AI, through the Microelectronics Commons (ME Commons) program of the U.S. Department of Defense (DoD), administered by the Applied Research Institute (ARI) and in part by the Center for the Co-Design of Cognitive Systems (CoCoSYS), a research center under the Joint University Microelectronics Program (JUMP) 2.0, a Semiconductor Research Corporation (SRC) initiative sponsored by DARPA.

\section{Declarations}
\subsection{Author Contributions}
A.H. and K.R. conceived the idea. A.H. directed the study, performed algorithmic studies and benchmarking, defined the macro design, and contributed to analog circuit design. S.C. designed and simulated the FSM control. S.S. contributed to circuit design and performed circuit-level simulations of components of the LIMO macro. A.H. and S.C. conducted data-analysis on the circuit-level results. A.M. conducted the cycle-accurate system-level simulations. F.G.R. developed the compact STT-MTJ model. D.B., F.I., and K.R. supervised and guided the study. A.H., K.R., and S.C. wrote the manuscript. All authors reviewed and approved the manuscript. 
\subsection{Competing Interests}
All authors declare no financial or non-financial competing interests

\subsection{Data Availability}
The TSPLib dataset used in this study is available at \url{http://comopt.ifi.uni-heidelberg.de/software/TSPLIB95/}. The CIFAR-10 dataset is available at \url{https://www.cs.toronto.edu/~kriz/cifar.html}, and the WIDER-FACE dataset is available at \url{http://shuoyang1213.me/WIDERFACE/}. The PASCAL VOC data used for testing face detection is available at \url{http://host.robots.ox.ac.uk/pascal/VOC/voc{year}/}, where \texttt{year} is $\{2007, 2008, 2009, 2010\}$.

\subsection{Code Availability}
The code for the digital twin of our architecture comprising multiple LIMO macros, used for qualitative benchmarking on large-scale TSPs, along with the code for hardware-aware CNN training, is publicly available at \url{https://github.com/a-holla/LIMO/tree/main}. The code for performing cycle-accurate simulations of our system following the PUMA architecture is available at \url{https://github.com/Aayush-Ankit/puma-simulator}.
The Verilog code used to synthesize the digital logic of the LIMO macro is not publicly available due to existing non-disclosure agreements. The Verilog-A compact model of the STT-MTJ is proprietary IMEC IP and is therefore not publicly available.


\bibliography{sn-bibliography}

\section{Figure Legends}
\begin{enumerate}
    \item (a) Illustration of the von Neumann bottleneck between memory and compute units for CO (b) Network-of-spins reformulation of COPs, with state access time growing quadratically. (c) A spatial architecture based on CiM primitives that need to efficiently solve divided CO instances. The considered architecture is organized hierarchically. Multiple tiles are connected by a network-on-chip. Each tile comprises multiple cores, and each core contains multiple compute units.
    \item (a) Navigation of the energy landscape in simulated annealing. Each valid tour of cities has an associated tour length, analogous to an energy state. The simulated annealing algorithm iteratively modifies the current tour, primarily accepting changes that decrease the tour length. To escape local minima, the algorithm occasionally accepts energy-increasing moves. The probability of accepting these moves diminishes over the course of the optimization. (b) Mapping of a TSP to $N^2$ spins. Each row represents a tour position and is a one-hot vector representing a city that is visited at that tour position. (c) A state transition within the network-of-spins representation for the TSP. To maintain a valid tour, a move consists of swapping the columns corresponding to two cities, which is analogous to swapping their visiting order.
    \item (a) Iterative refinement process in Metropolis-SA. An edge heat-map is shown for a 7-city TSP instance that denotes the probability of accepting a swap between two cities if the pair is selected. Only edges with $P_{\text{accept}}>0.3$ are shown for clarity. (b) An illustration of the SWAI algorithm, that relies on iterative tour construction. A node heat-map is shown that denotes the selection probability. (c) Comparison of Metropolis-SA and SWAI on 500 random TSP instances per problem size, ranging from $N=9$ to $N=24$. Inter-quartile ranges are shaded around the median line-plots.
    \item Applying hardware related constraints to the SWAI algorithm. Inter-quartile ranges are shaded around the median line-plots. (a) Varying the bit-width of the weight matrix. (b) Varying the number of passes of the algorithm. (c) Varying the number of restart runs for a given problem instance.
    \item Floorplan of the proposed LIMO macro. At its core is an 80$\times$80 8T-SRAM CiM crossbar that doubles as (i) the spin array with 4-bit programmable couplings for Ising annealing and (ii) a weight stationary matrix for neural-network inference. Distinct peripheral blocks are provisioned for the two operating modes; the annealing peripheral incorporates a compact STT-MRAM TRNG module. Five independent TSPs are concurrently mapped onto the crossbar and solved in parallel, while a scratch SRAM records the best tour for each instance. A finite-state-machine (FSM) based control module coordinates all dataflows between macro components across both modes. (a) Elementary bit-cell operations supported by the compute macro, including standard row-reads and ternary weight multiplication. Ternary weights are realized by a push-pull design in the read paths of neighboring bit-cells, which facilitates a bi-directional current flow to the source line. (b) Dataflow in annealing mode. (d) Dataflow in VMM mode.
    \item (a) Illustrative example of the threshold based comparison to generate a Bernoulli sampler with a $3/8$ success probability from a 3-bit uniformly distributed number. This same concept is used to generate a stochastic gating vector that is dependent on the coupling constant value. (b) Ideal geometric decay with $p_0=0.2$ and $\beta=0.9995$ and its piecewise linear fit. (c) Circuit schematic of the sense‑amplifier‑based STT‑MRAM TRNG (d) XOR combination of the bitstreams from two identical instances yields a composite random sequence that is robust against device-to-device variations \cite{rhs-trng}. (e) Switching probability for the $R_{AP}\rightarrow R_{P}$ and $R_{P}\rightarrow R_{AP}$  transitions as a function of drive current (10ns pulse width) under fast, slow, and typical corners. The inset on the right plots the corresponding read resistance of the MTJ versus bias voltage for the same corners.
    \item Various steps involved in our divide-and-conquer framework for solving large scale TSPs. (a) Bottom-up clustering process and top-down solving process for two hierarchical levels. (b) Tour stitching process for the meta-centroids/nodes at level $i-1$, based on the TSP ordering of centroids in level $i$. (c) Parallelized segment refinement process at level $i$. (d) 2-opt local refinement process at level $i$.
    \item Breakdown of power consumption among the LIMO macro's components when operating in (a) annealing mode and (b) VMM mode at 100MHz.
    \item A comparison of deviation ratios for various TSP-annealers in literature. The deviation is measured relative to optimal solutions from exact solvers, with lower values indicating superior performance. The numerical value in each problem ID denotes the number of nodes of the TSP.
    \item (a) TSP Solver runtime comparison: LIMO vs. TAXI vs. NeuroIsing. (b) Runtime profiling of the LIMO algorithm, showing the percentage contribution of each constituent step for the same problem instances.
    \item (a) Comparison of normalized latency and energy consumption between our effectively ADC-less LIMO macro and a conventional CiM array that utilizes an energy-efficient SAR-ADC. Results are shown for the two types of CNNs considered: Resnet20 (for CIFAR-10 image classification) and ResnetSSD (for face detection). (b) Breakdown of energy consumption by component for the PUMA-based spatial architecture incorporating our macro, for the two CNNs.
    \item Characteristics of the MTJ as predicted by the compact model \cite{vlsisttdevice2025}, which has been calibrated to experimental data. (a, b) Nominal (TT corner) switching probabilities for the $R_{AP} \rightarrow R_{P}$ and $R_{P} \rightarrow R_{AP}$
  transitions, respectively, plotted as a function of write current amplitude and pulse width. (c) Nominal resistance values for the two MTJ states, as determined by the model's calibration to experimental R-V characteristics.
\end{enumerate}
\end{document}


\begin{center}
    {\LARGE \textbf{Supplementary Information}}\\[1em]
    {\large \textbf{LIMO: \underline{L}ow-Power \underline{I}n-Memory-Annealer and \underline{M}atrix-Multiplication Primitive for Edge C\underline{o}mputing}}\\[1em]
    
    \textbf{Amod Holla}$^{1,\dagger}$, 
    \textbf{Sumedh Chatterjee}$^{2,*}$, 
    \textbf{Sutanu Sen}$^{3,*}$, 
    \textbf{Anushka Mukherjee}$^1$,\\
    \textbf{Fernando Garc\'{i}a-Redondo} $^4$,
    \textbf{Dwaipayan Biswas}$^4$, 
    \textbf{Francesca Iacopi}$^{1,5}$, 
    \textbf{Kaushik Roy}$^{1,\dagger}$\\[1em]

    $^1$Elmore School of Electrical and Computer Engineering, Purdue University, West Lafayette, IN, USA\\
    $^2$Department of EE, IIT Madras, Chennai, India\\
    $^3$Department of ECE, IIT Kharagpur, Kharagpur, India\\
    $^4$imec, Leuven, Belgium\\
    $^5$imec USA, West Lafayette, IN, USA\\[0.5em]
    
    *These authors contributed equally.\\
    $\dagger$Corresponding authors: \texttt{\{aholla, kaushik\}@purdue.edu}
\end{center}


\section{Control and Driving Logic}

\subsection{The Control FSM}

The FSM responsible for executing the annealed insertions and VMMs in the LIMO macro transitions through the following sequence of operational states, and as outlined in table \ref{fsm_table}.

\begin{itemize}
    \item \textbf{PRG\_ROW}: Initializes the macro by programming the weight matrix and initial spin configuration onto the crossbar array. Also programs a weight-stationary matrix for the VMM mode.

    \item \textbf{GEN}: Asserts the global enable signal and generates a global 16-bit reference word to be broadcast to the global TRNG.

    \item \textbf{LEN}: Asserts the local enable signal and holds it active for the duration required to generate local 4-bit stochastic words for each city (in order to follow the local-stochasticity scheme described in the main manuscript).

    \item \textbf{SS\_RD}: Accesses the spin-storage partition at the index corresponding to the city visited immediately prior to the current one. The final two clock cycles of the local enable signal are pipelined in this state to facilitate readout of the previous city's spin state.

    \item \textbf{W\_RD}: Retrieves the distance weights (or coupling constants) from the previously visited city by reading the crossbar outputs. After shift-and-add, the distances are passed through local TRNG units and a comparator tree for stochastic evaluation.

    \item \textbf{STO\_SOLN}: Writes the comparator tree output to the spin-storage array at the current index. It also performs candidate elimination, computes a running sum of selected weights in real-time, and stores intermediate results in a scratchpad 6T-SRAM array.

    \item \textbf{LAST\_CITY}: Resets the registers used for tracking the running sum and candidate city list, preparing the system for the next traversal or problem instance.

    \item \textbf{AI\_RD} : Performs the VMM operation.
\end{itemize}

\begin{table}[t]
\centering
\caption{State--transition table of the LIMO control FSM}
\label{tab:fsm}
\renewcommand{\arraystretch}{1.15}
\begin{tabular}{llll}
\toprule
\textbf{Current state} & \textbf{Duration (clk)} & \textbf{Transition condition} & \textbf{Next state} \\
\midrule
\texttt{PRG\_ROW}   & 80 & last  index $\And$ ai mode                   & \texttt{AI\_RD} \\
                    &    & last  index $\And$ ising mode                   & \texttt{GEN} \\[2pt]

\texttt{GEN}        &  1 & $g\_bit = 1$ \footnotemark[1]                                                   & \texttt{LEN} \\
                    &    & $g\_bit = 0$  \footnotemark[1]                                                 & \texttt{SS\_RD} \\[2pt]

\texttt{LEN}        &  6 & always (after six cycles)                                              & \texttt{SS\_RD} \\[2pt]

\texttt{SS\_RD}     &  2 & always (after two cycles)                                      & \texttt{W\_RD} \\[2pt]

\texttt{W\_RD}      &  1 & always (after one cycle)                                       & \texttt{STO\_SOLN} \\[2pt]

\texttt{STO\_SOLN}  &  2 & last index                                                & \texttt{LAST\_CITY} \\
                    &    & not last index $\And$ last problem  $\And g\_bit = 1$
                                                                                          & \texttt{LEN} \\
                    &    & otherwise                                                      & \texttt{SS\_RD} \\[2pt]

\texttt{LAST\_CITY} &  1 & last problem $\And$ last pass
                                                                                          & \texttt{IDLE} \\
                    &    & last problem $\And$ not last pass
                                                                                          & \texttt{GEN} \\
                    &    & otherwise                                                      & \texttt{SS\_RD} \\[2pt]

\texttt{AI\_RD}     &  1 & always (after one cycle)                                       & \texttt{IDLE} \\[2pt]

\texttt{IDLE}       &  1 & external \textit{start} asserted                               & \texttt{PRG\_ROW} \\
\bottomrule

\end{tabular}
\footnotetext[1] {\textit{g\_bit} is the bit that arises from the comparision of the 16-bit global word with Global TRNG output.}
\label{fsm_table}
\end{table}

The looping logic maintained by the \texttt{pass} , \texttt{index}, and \texttt{problem} registers in anneal mode is given in algorithm \ref{alg:S1}, where a \texttt{pass} refers to a complete pass over the macro across all problems, \texttt{index} refers to the tour position ($2-16$) the macro is currently solving for one of the problems, and \texttt{problem} refers to which problem in the macro the solver is currently on ($1-5$).

\begin{algorithm}[H]
\caption{PASS, INDEX, and PROBLEM loop} \label{alg:S1}
\begin{algorithmic}[1]
\For{pass $\gets 1$ \textbf{to} $\textit{PASS\_COUNT}$}                     
    \If{\textit{open}}                                                      
        \State $index_{\max} \gets \textit{INDEX\_COUNT} - 1$
    \Else                                                                   
        \State $index_{\max} \gets \textit{INDEX\_COUNT}$
    \EndIf
    \For{index $\gets 2$ \textbf{to} $index_{\max}$}                        
        \If{$index = 1$}                                                    
            \State \Call{GEN}{}
        \EndIf

        \For{problem $\gets 1$ \textbf{to} $\textit{PROBLEM\_COUNT}$}       
            \If{$problem = 0$}                                              
                \If{$g\_bit = 1$}
                    \State \Call{LEN}{}
                \EndIf
            \EndIf

            \State \Call{SS\_RD}{index{-}1,\;problem}
            \State \Call{W\_RD}{problem}
            \State \Call{STO\_SOLN}{index,\;problem}
        \EndFor

        \If{\textbf{last index}}                                            
            \State \Call{LAST\_CITY}{}
        \EndIf
    \EndFor
\EndFor
\end{algorithmic}
\end{algorithm}

\subsection{Global TRNG Reference Generation and Enable Signals}

\begin{itemize}
    \item \textbf{Global TRNG Reference Generation}: This is generated by initializing a 16-bit word that provides the initial stochasticity, followed by a decrement at the start of every pass during the GEN state. The decrement value can be picked from a suite of available slopes.The slopes enable the 16-bit word to decay in a piece-wise fashion that resembles exponential schedules  The LIMO macro can support the decay annealing schedules 0.9995 and 0.995 through piece-wise linear fitting, with the slopes shown in table \ref{slope_suite}. In the event the value in the word register is less than the decrement value, the annealing process is completed. \\

    \item \textbf{Enable signals}: The enable to the global TRNG is asserted during the GEN state, whereas the enable to the local TRNG is asserted during the LEN state , as well as the SS\_RD, conditional to the \textit{g\_bit}.
\end{itemize}

\begin{table}[]
\centering
\caption{Piece-wise constant decrement schedule.}
\begin{tabular}{@{}ccccc@{}}
\toprule
\multicolumn{2}{c}{\textbf{0.9995 decay rate}} & & \multicolumn{2}{c}{\textbf{0.9950 decay rate}} \\[-0.2em]
\cmidrule(lr){1-2}\cmidrule(l){4-5}
\textbf{Slope} & \textbf{Interval (passes)} & & \textbf{Slope} & \textbf{Interval (passes)} \\ \midrule
10 & $0 \to 267$          & & 10 & $0 \to 27$ \\
 8 & $267 \to 575$        & &  8 & $27 \to 57$ \\
 7 & $575 \to 940$        & & 7 & $57 \to 94$ \\
 5 & $940 \to 1\,386$     & & 5 & $94 \to 138$ \\
 4 & $1\,386 \to 1\,961$  & &  4 & $138 \to 196$ \\
 3 & $1\,961 \to 2\,772$  & &  3 & $196 \to 277$ \\
 2 & $2\,772 \to 4\,158$  & &  2 & $277 \to 358$ \\
 1 & $4\,158 \to 5\,990$  & &   &  \\ \bottomrule
\end{tabular}
\label{slope_suite}
\end{table}

\begin{figure}[h]
\centering
\includegraphics[width=\textwidth]{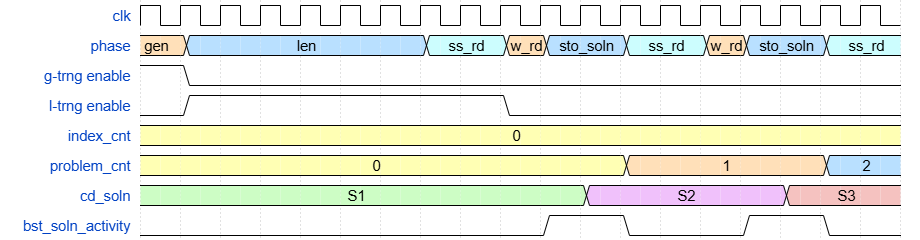}
\caption{System Timing Diagram for Annealing Mode.}\label{fig_s5}
\end{figure}

\subsection{Stage-wise Drivers}

Instead of a single general‐purpose row-column decoder, the design employs distinct decoders for each state. Each state‐specific decoder is enabled only during its corresponding phase, and their outputs are merged by a top‐level multiplexer, making the operations faster. \\

\begin{itemize}

  \item \textbf{PRG\_ROW} : The \emph{write-word-line} (WWL) is activate one row at a time, and all the weight and spin storage values corresponding to that index is written from the bit-lines. \\
  \item \textbf{SS\_RD}:  
        Drives the \emph{read-word-lines } (RWL and RWLB) of the \emph{spin-storage} partition at \mbox{index – 1}.
        This readback exposes the city chosen in that earlier step. \\

  \item \textbf{W\_RD}:  
        Activates the RWL and RWLB of the \emph{weight} partitions in the row corresponding to the city returned by \textbf{SS\_RD}.  
        The crossbar therefore produces the relevant distances, which are forwarded to the peripheral circuitry for further processing.\\

  \item \textbf{STO\_SOLN}:  
        Two operations are executed in a pipelined fashion.  
        \begin{enumerate}
          \item The newly selected city (available at the comparator-tree output) is written back to the SS partition.  
          \item Concurrently, the edge-weight between the city at \mbox{index – 1} and the new city is computed by asserting the RWL/RWLB of the chosen-city row \emph{and} the bit-lines that encode the current index.  
                The resulting signal is sensed by the sense amplifier and post-processed.  
        \end{enumerate}
        This concurrency is possible because, in 8T-SRAM, RWLs and WWLs are decoupled. \\

  \item \textbf{AI\_RD}:  
        Implements the vector–matrix-multiplication (VMM) step.  
        The input vector is applied to the RWLs and RWLBs, while the bit-lines are alternatively asserted according to the scheme described in the main text.
\end{itemize}

\subsection{Storing the Best Solution}

During parameter‐space exploration, the macro continuously computes the sum metric corresponding to the total tour distance for each candidate solution. At the completion of each pass, when the solver reaches the final city index, the current sum register is compared against the stored best‐sum register. If the current sum exceeds the previously recorded best, the best‐sum register is updated with the new value, and the \texttt{parity} bit is toggled. There are 5 such register banks and \texttt{parity} bits (described subsequently), corresponding to the 5 parallel TSP problems that can be solved on the macro. The current sum registers are reset during the LAST\_CITY states. \\ 

For storing the best-solutions for all 5 problems, a 6T‐SRAM scratch array is employed. Each problem instance is allocated two sub‐arrays, one per parity state. During each STO\_SOLN stage, the comparator‐tree output is written into the active sub‐array row. When the \texttt{parity} bit flips, subsequent writes are directed to the alternate sub‐array, thereby preserving the newly discovered best solution.

\subsection{Candidate Masking and Update}

During each pass, the solver must avoid re‐selecting cities that have already been visited. To enforce this, the local TRNG output is AND-ed with a \texttt{candidate} register, which acts as a bitmask of available cities. After every STO\_SOLN stage, the bit corresponding to the newly chosen city in the active \texttt{candidate} register is cleared (from 1 to 0), preventing its reselection in subsequent steps. The macro instantiates five independent \texttt{candidate} registers--one for each concurrent TSP problem. \\ In open‐loop mode, both the initial and terminal cities are preemptively removed from the candidate mask to prevent their selection. In closed‐loop mode, only the designated start city is excluded by default. Upon entering the LAST\_CITY state, all exclusions are cleared, reinstating every city into the candidate set for the next pass.

\subsection{Miscellaneous Power Saving Techniques}

\begin{itemize}
  \item \textbf{Clock Gating}:  
        Integrated Clock Gating Cells are instantiated via the Genus Synthesis Solution.  
        All multi‐bit registers are gated with enable signals so that they receive the clock only when an update is required, thereby reducing power. There is an increase in clock power but this is small compared to the power saved by clock-gating as a whole.

  \item \textbf{Operand Isolation}:  
        To suppress unnecessary switching in large combinational driver trees, operand‐isolation logic is inserted.  
        Simple \texttt{AND} gates and multiplexers block inactive inputs, preventing toggling and lowering power consumption.
\end{itemize}

\section{Analog Peripherals, Layouts and Area Statistics}
\subsection{STT-MTJ Based TRNG}
\begin{figure}[h]
\centering
\includegraphics[width=\textwidth]{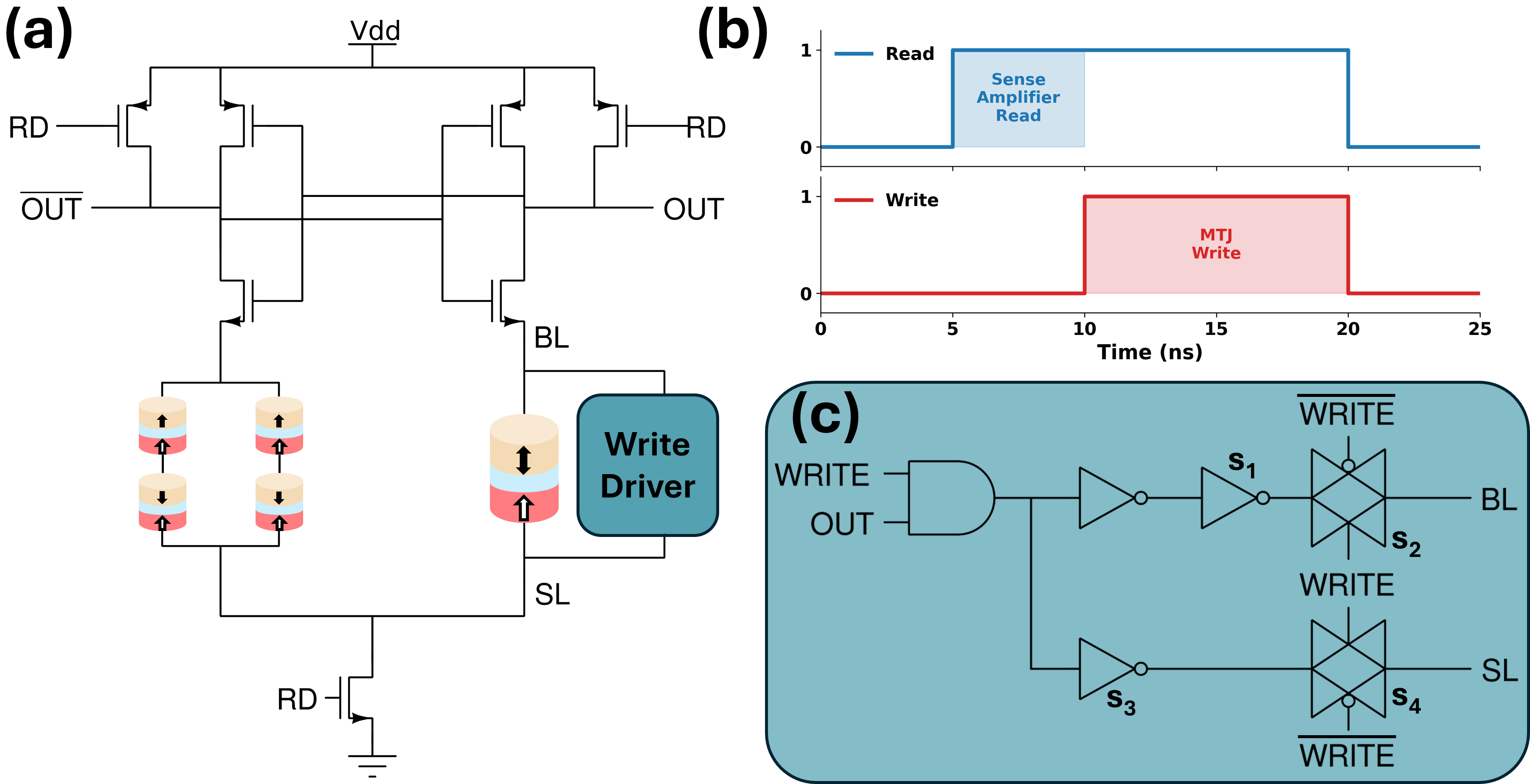}
\caption{(a) Schematic of the sense-amplifier based TRNG. (b) Read and Write waveforms during TRNG operation. (c) Schematic of bi-directional write-driver.}\label{fig_s2}
\end{figure}

\begin{figure}[h]
\centering
\includegraphics[width=0.78\textwidth]{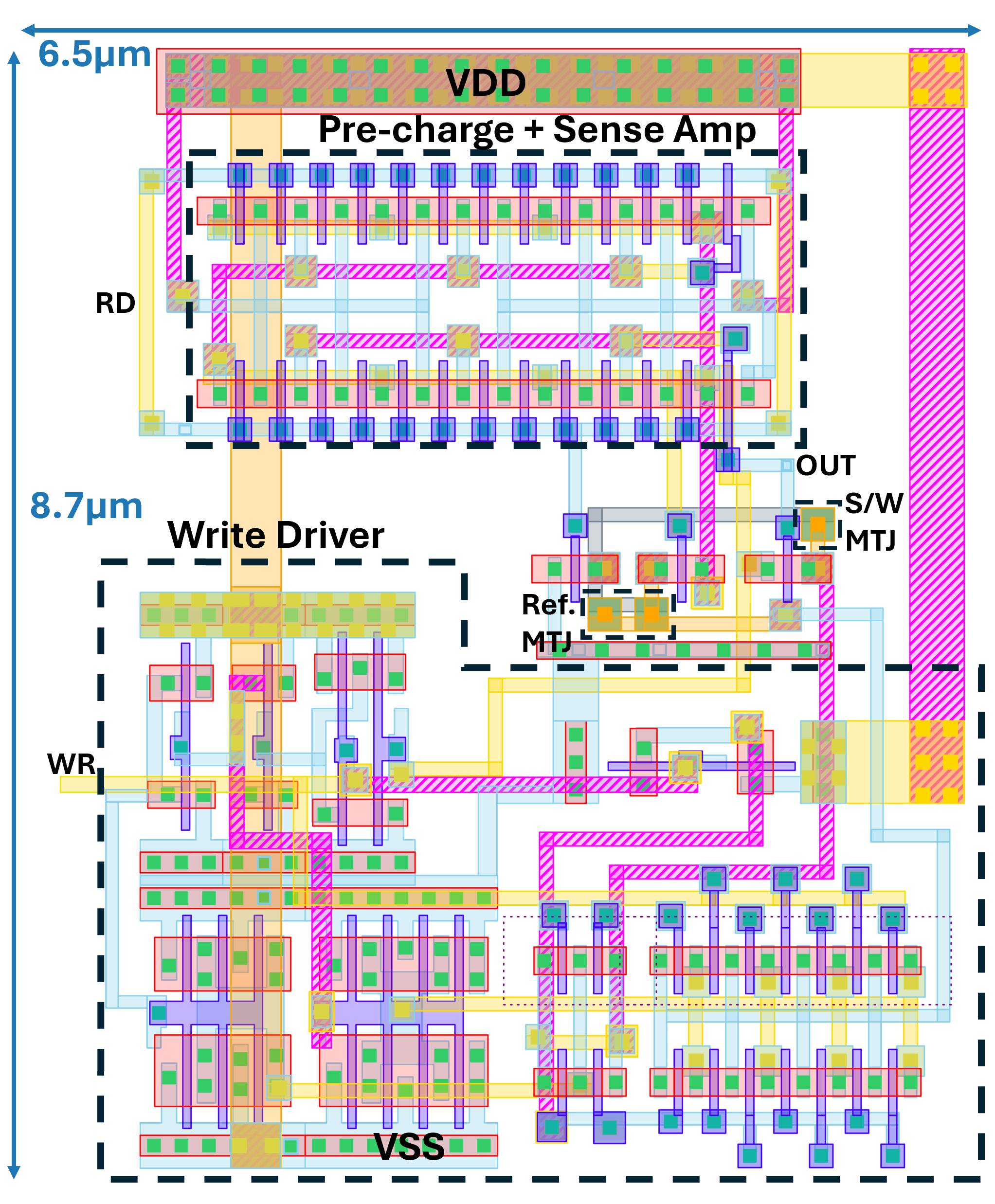}
\caption{Layout of the STT-MTJ TRNG used in the LIMO macro.}\label{fig_s3}
\end{figure}

Fig. \ref{fig_s2}(a) shows the differential STT-MTJ based sense-amplifier unit that goes into the LIMO macro \cite{mtjsa}. It amplifies the difference between the resistance of the two branches. The right-MTJ is switchable and determines the output of the TRNG. The left-branch is a mid-point reference and comprises of two fixed-state MTJs forming an equivalent resistance of $(R_{P}+R_{AP})/2$. The operational cycle, depicted in Figure~\ref{fig_s2}(b), is divided into three distinct phases over a 20 ns period. The cycle begins with a 5ns precharge stage. This is followed by a 5ns read phase, during which the sense amplifier resolves the switchable-MTJ state to a digital output. The final 10ns are dedicated to a bidirectional write phase, where the right-MTJ is stochastically switched based on the result of the preceding read. The write driver \cite{rhs-trng}, depicted in Figure~\ref{fig_s2}(c), performs the state-dependent, bi-directional switching of the right-MTJ. The inverters and pass gates are appropriately sized to push a write current that induces a 50\% switching probability in either direction. The layout of the TRNG is shown in Fig. \ref{fig_s3}. 

\FloatBarrier
\newpage
\subsection{Sense Amplifier}
\begin{figure}[h]
\centering
\includegraphics[width=\textwidth]{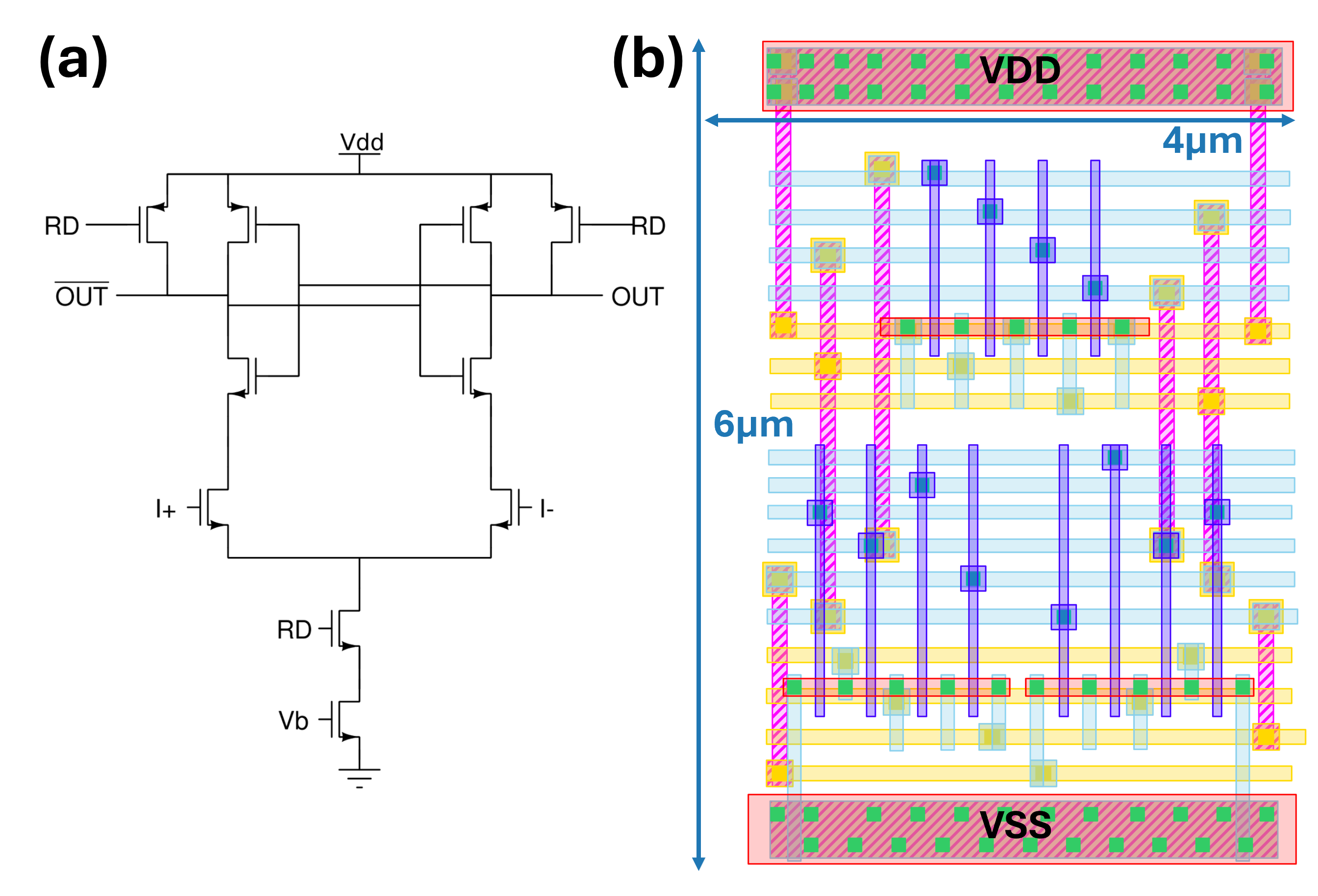}
\caption{(a) Schematic and (b) Layout of the sense amplifier used in the LIMO macro.}\label{fig_s4}
\end{figure}
\subsection{Area Breakdown}
\begin{figure}[h]
\centering
\includegraphics[width=0.79\textwidth]{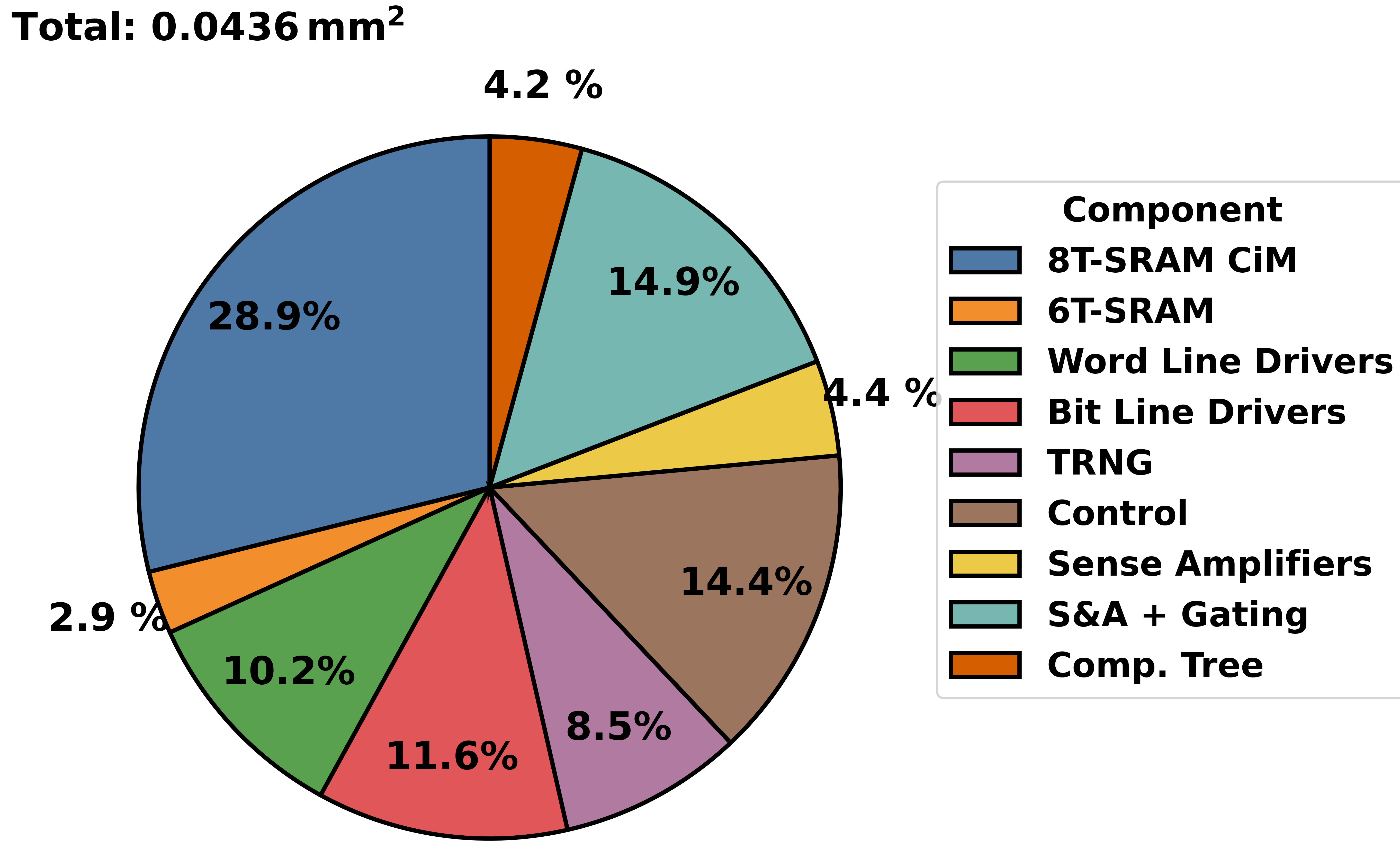}
\caption{Contributions of various components to the total area of the LIMO macro.}\label{fig_s6}
\end{figure}
\section{PCA-Based Clustering}
\begin{figure}[h]
\centering
\includegraphics[width=0.8\textwidth]{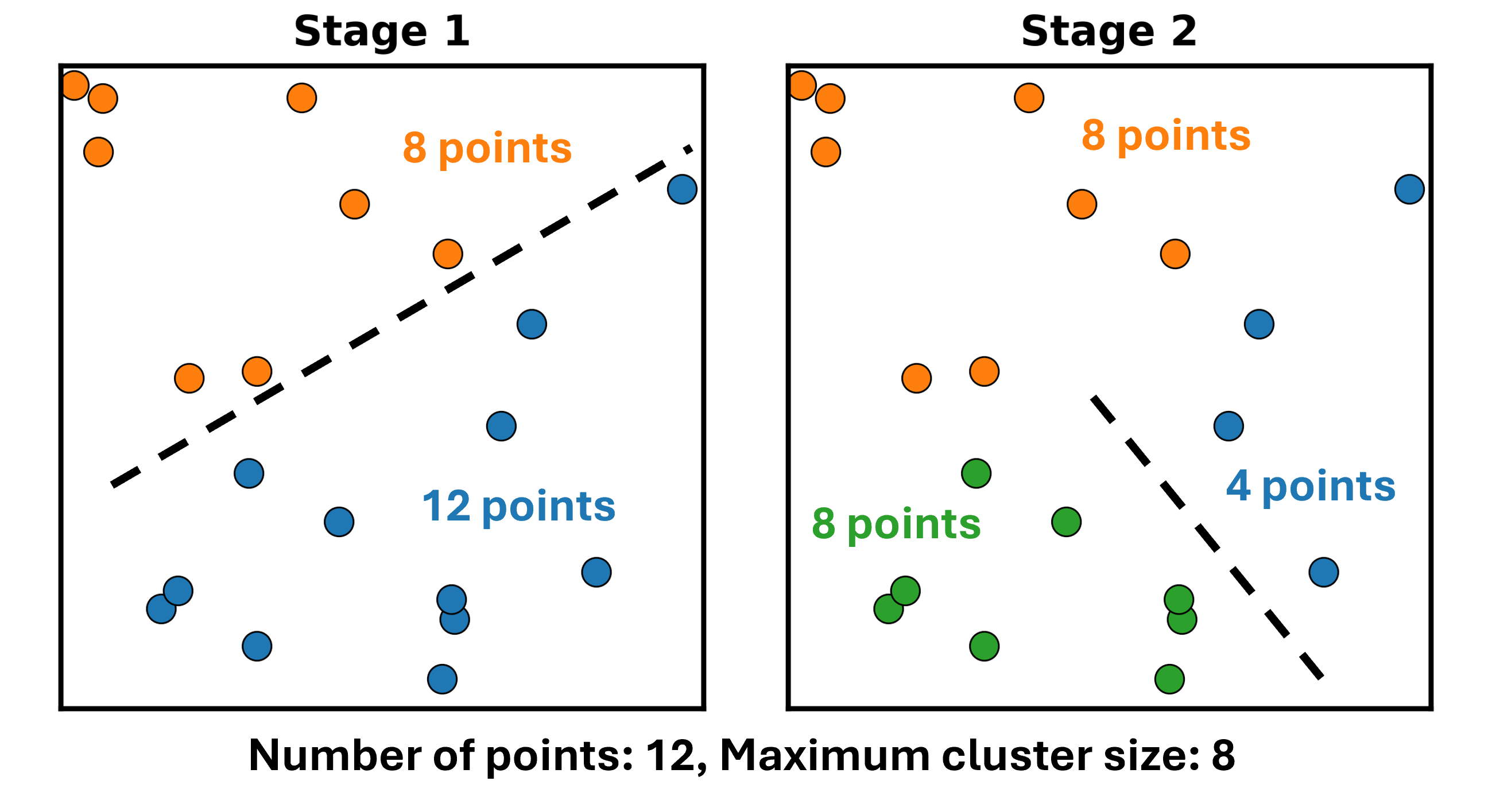}
\caption{Illustrative example of the clustering algorithm on 20 points, with a maximum cluster size of 8.}\label{fig_s1}
\end{figure}

The clustering algorithm adopts a projection-based strategy \cite{Boley1998PDDP} that repeatedly bisects the data until every terminal cluster contains at most \(M\) points. For a subset \(S=\{\mathbf{x}_i\}_{i=1}^{n}\subset\mathbb{R}^{2}\), it first evaluates the mean
\[
\boldsymbol{\mu}=
\begin{pmatrix}
\mu_x\\[2pt]
\mu_y
\end{pmatrix},
\qquad
\mu_x=\frac{1}{n}\sum_{i=1}^{n}x_i,
\quad
\mu_y=\frac{1}{n}\sum_{i=1}^{n}y_i,
\]
and the unbiased covariance matrix
\[
\boldsymbol{\Sigma}=
\frac{1}{n-1}\sum_{i=1}^{n}
(\mathbf{x}_i-\boldsymbol{\mu})
(\mathbf{x}_i-\boldsymbol{\mu})^{\!\top}.
\]
The dominant eigenvector \(\mathbf{v}\) of \(\boldsymbol{\Sigma}\) defines the axis of maximal variance. Each point is then projected as \(s_i=\mathbf{v}^{\top}\mathbf{x}_i\), and the ordered projections \(\{s_{(i)}\}_{i=1}^{n}\) are used to identify a cut index \(k\) that maximizes the between-cluster variance \cite{Otsu1979Threshold}:
\[
V_{\mathrm{B}}(k)=
k\bigl(\bar{s}_1-\bar{s}\bigr)^{2}
+
(n-k)\bigl(\bar{s}_2-\bar{s}\bigr)^{2},
\]
where
\(\bar{s}_1=\frac{1}{k}\sum_{i=1}^{k}s_{(i)}\),
\(\bar{s}_2=\frac{1}{n-k}\sum_{i=k+1}^{n}s_{(i)}\), and
\(\bar{s}=\frac{1}{n}\sum_{i=1}^{n}s_i\).
The optimal split
\(k^{*}=\arg\max_{1\le k<n}V_{\mathrm{B}}(k)\)
partitions \(S\) into two daughter clusters on which the procedure recurses, yielding a binary dendrogram whose leaves satisfy \(|S|\le M\).
By aligning each cut with the principal component and selecting the break that maximizes \(V_{\mathrm{B}}\), the method effectively maximizes a first-order Fisher discriminant ratio at every stage, producing relatively compact and well-separated clusters. This process is illustrated in Fig. \ref{fig_s1}.
\newpage
\section{Divide-and-Conquer for Large TSPs -- Ablations}
\begin{table}[h]
  \caption{Ablation deviation ratios and relative changes from baseline for the two algorithmic steps.}\label{tab:ablation_ratios}
  \centering
  \begin{tabular}{@{}l c cc cc@{}}
    \toprule
    Problem & Baseline & \multicolumn{2}{c}{Without 2-opt} & \multicolumn{2}{c}{Without Segment Refinement} \\
    \cmidrule(lr){3-4} \cmidrule(lr){5-6}
            &          & Deviation Ratio & \% Change v/s Baseline & Deviation Ratio & \% Change v/s Baseline \\
    \midrule
    \texttt{kroE100}   & 1.024197934 & 1.054830524 & +2.99\%  & 1.048214609 & +2.34\% \\
    \texttt{kroB200}   & 1.035454948 & 1.038966352 & +0.34\%  & 1.063034807 & +2.66\% \\
    \texttt{gil262}    & 1.061842659 & 1.126209508 & +6.06\%  & 1.055111485 & -0.63\% \\
    \texttt{lin318}    & 1.047675805 & 1.109797378 & +5.93\%  & 1.068724672 & +2.01\% \\
    \texttt{pcb442}    & 1.063447558 & 1.100939296 & +3.53\%  & 1.087861249 & +2.30\% \\
    \texttt{rat575}    & 1.066351411 & 1.097384365 & +2.91\%  & 1.097088813 & +2.88\% \\
    \texttt{gr666}     & 1.068166555 & 1.140698455 & +6.79\%  & 1.120214909 & +4.87\% \\
    \texttt{rat783}    & 1.060413355 & 1.090847150 & +2.87\%  & 1.101181013 & +3.84\% \\
    \texttt{pr1002}    & 1.068856954 & 1.113197099 & +4.15\%  & 1.129016268 & +5.63\% \\
    \texttt{u1060}     & 1.072585611 & 1.109628102 & +3.45\%  & 1.131788446 & +5.52\% \\
    \texttt{d2103}     & 1.131843645 & 1.180443172 & +4.29\%  & 1.245051662 & +10.00\% \\
    \texttt{u2152}     & 1.142545873 & 1.182092665 & +3.46\%  & 1.264532395 & +10.68\% \\
    \texttt{pr2392}    & 1.081262962 & 1.114485017 & +3.07\%  & 1.188690375 & +9.94\% \\
    \texttt{pcb3038}   & 1.112401412 & 1.130840850 & +1.66\%  & 1.202049472 & +8.06\% \\
    \texttt{fnl4461}   & 1.096162484 & 1.112348411 & +1.48\%  & 1.180438855 & +7.69\% \\
    \texttt{rl5915}    & 1.194219582 & 1.227469807 & +2.78\%  & 1.355280887 & +13.49\% \\
    \texttt{rl5934}    & 1.176390400 & 1.208270913 & +2.71\%  & 1.319154025 & +12.14\% \\
    \texttt{rl11849}   & 1.162705875 & 1.185229198 & +1.94\%  & 1.314283331 & +13.04\% \\
    \texttt{d18512}    & 1.120943989 & 1.129747385 & +0.79\%  & 1.232800903 & +9.98\% \\
    \texttt{pla33810}  & 1.166757516 & 1.187769106 & +1.80\%  & 1.305201038 & +11.87\% \\
    \texttt{pla85900}  & 1.153408295 & 1.163318701 & +0.86\%  & 1.281159077 & +11.08\% \\
    \botrule
  \end{tabular}
\end{table}
\FloatBarrier
\bibliography{sn-bibliography}